\documentclass[aps,onecolumn,floats,prl]{revtex4}

\usepackage{changepage}

\usepackage[utf8]{inputenc}

\usepackage{textcomp,marvosym}

\usepackage{amsmath,amssymb}

\usepackage{nameref,hyperref}

\usepackage{rotating}

\begin{document}

\title{Dynamics of transcription factor binding site evolution}

\author{Murat Tu\u{g}rul\footnote{mtugrul@ist.ac.at}, Tiago Paix\~{a}o, Nicholas H. Barton,  Ga\v{s}per Tka\v{c}ik
}

\affiliation{Institute of Science and Technology Austria, Am Campus 1, A-3400 Klosterneuburg, Austria}

\date{\today}

\begin{abstract}
Evolution of gene regulation is crucial for our understanding of the phenotypic differences between species, populations and individuals. Sequence-specific binding of transcription factors to the regulatory regions on the DNA is a key regulatory mechanism that determines gene expression and hence heritable phenotypic variation. We use a biophysical model for directional selection on gene expression to estimate the rates of gain and loss of transcription factor binding sites (TFBS) in finite populations under both point and insertion/deletion mutations. Our results show that these rates are typically slow for a single TFBS in an isolated DNA region, unless the selection is extremely strong. These rates decrease drastically with increasing TFBS length or increasingly specific protein-DNA interactions, making the evolution of sites longer than $\sim 10$ bp unlikely on typical eukaryotic speciation timescales. Similarly, evolution converges to the stationary distribution of binding sequences very slowly, making the equilibrium assumption questionable. The availability of longer regulatory sequences in which multiple binding sites can evolve simultaneously, the presence of  ``pre-sites'' or partially decayed old sites in the initial sequence, and biophysical cooperativity between transcription factors, can all facilitate gain of TFBS and reconcile theoretical calculations with timescales inferred from comparative genomics. 
\end{abstract}

\maketitle

\section*{Author Summary}
Evolution has produced a remarkable diversity of living forms that manifests in qualitative differences as well as quantitative traits.
An essential factor that underlies this variability is transcription factor binding sites, short pieces of DNA  that control gene expression levels. Nevertheless, we lack a thorough theoretical understanding of the evolutionary times required for the appearance and disappearance of these sites. By combining a biophysically realistic model for how cells read out information in transcription factor binding sites with model for DNA sequence evolution, we explore these timescales and ask what factors crucially affect them. We find that the emergence of binding sites from a random sequence is generically slow under point and insertion/deletion mutational mechanisms. Strong selection, sufficient genomic sequence in which the sites can evolve, the existence of partially decayed old binding sites in the sequence, as well as certain biophysical mechanisms such as cooperativity, can accelerate the binding site gain times and make them consistent with the timescales suggested by comparative analyses of genomic data.


\section{Introduction}

Evolution produces heritable phenotypic variation within and between populations and species on relatively short timescales. 
Part of this variation is due to differences in gene regulation, which determines how much of each gene product exists in every cell. 
These gene expression levels are heritable quantitative traits subject to natural selection~\cite{fay_evaluating_2007, zheng_regulatory_2011, romero_comparative_2012}.
While the importance of their variability for the observed phenotypic variation is still debated~\cite{hoekstra_locus_2007}, it is believed to be crucial within closely related species or in populations whose proteins are functionally or structurally similar~\cite{wittkopp_evolution_2013}. 
The genetic basis for gene expression differences is thought to be non-coding regulatory DNA, but our understanding of its evolution is still immature;  this is due, in part, to the lack of precise knowledge about the mapping between the regulatory sequence and the resulting expression levels. 

Transcriptional regulation is the most extensively studied mechanism of gene regulation. 
Transcription factor proteins (TFs) recognize and bind specific DNA sequences called binding sites, thereby affecting the expression  of target genes.
Eukaryotic regulatory sequences, i.e., enhancers and promoters, are typically between a hundred and several thousand base pairs (bp) in length~\cite{yao_coexpression_2015}, and can harbor many transcription factor binding sites (TFBSs), each typically consisting of $6-12$ bp. 
The situation is different in prokaryotes: they lack enhancer regions and have one or a few TFBSs which are typically
longer, between 10 to 20 bp in length \cite{wunderlich_different_2009, stewart_why_2012}. Differences in TF binding are thought to arise primarily due to changes in the regulatory sequence at the TF binding sites rather than changes in the cellular environment or the TF proteins themselves~\cite{schmidt_five-vertebrate_2010}. Nevertheless, a theoretical understanding of the relationship between the evolution of the regulatory sequence and the evolution of gene expression levels remains elusive, mostly because of the complex interaction of evolutionary forces and biophysical processes~\cite{stefflova_cooperativity_2013}.

From the evolutionary perspective, the crucial question is whether and when these regulatory sequences can evolve rapidly enough so that new phenotypic variants can arise and fix in the population over typical speciation timescales.
Comparative genomic studies in eukaryotes provide evidence for the evolutionary dynamics of TF binding, highlighting the possibility for rapid and flexible TFBS gain and loss between closely related species on timescales of as little as a few million years~\cite{dowell_transcription_2010, villar_evolution_2014}.
Examples include quick gain and loss events that cause divergent gene expression~\cite{doniger_frequent_2007}, or the compensation of such events by turn-over at other genome locations~\cite{moses_large-scale_2006}; gain and loss events sometimes occur even in the presence of strong constraints on expression levels~\cite{ludwig_functional_1998, paris_extensive_2013}. Furthermore, such events enabled new binding sites on sex chromosomes that arose as recently as $1-2$ million years ago~\cite{ellison_dosage_2013, alekseyenko_conservation_2013}. There are  examples of rapid regulatory DNA evolution across and within populations requiring shorter timescales, i.e. $10.000-100.000$ years~\cite{contente_polymorphic_2002, kasowski_variation_2010, zheng_regulatory_2011,chan_adaptive_2010}. 
On the other hand, strict conservation has also been observed at orthologous regulatory locations even in distant species (e.g., \cite{vierstra_mouse_2014}). 
Taken together, these facts suggest that the rates of TFBS evolution can extend over many orders of magnitude and differ greatly from the point mutation rate at a neutral site. 
To study the evolutionary dynamics of regulatory sequences and understand the relevant timescales, we set up a theoretical framework with a special focus on the interplay of both population genetic and biophysical factors, briefly outlined below. 

Sequence innovations originate from diverse mutational mechanisms in the genome. 
While tandem repeats~\cite{gemayel_variable_2010} or transposable elements~\cite{feschotte_transposable_2008} may be important in evolution, the better studied and more widespread mutation types still need to be better understood in the context of TFBS evolution. 
Specifically, we ask how the evolutionary dynamics are affected by single nucleotide (point) mutations, as well as by insertions and deletions (indels).
New mutations in the population are selected or eliminated by the combined effects of selection and random genetic drift. 
Although the importance of selection~\cite{hahn_effects_2003,he_does_2011,arnold_quantitative_2014} and mutational closeness of the initial sequences~ \cite{macarthur_expected_2004, nourmohammad_formation_2011} for TF binding site evolution has already been reported, the belief in fast evolution via point mutations without selection (i.e., neutral evolution) persists in the literature (e.g.,\cite{wittkopp_evolution_2013, villar_evolution_2014}), mainly due to Stone \& Wray's (2001) misinterpretation of their own simulation results~\cite{stone_rapid_2001} (see Macarthur \& Brookfield (2004)~\cite{macarthur_expected_2004}). 
This likely reflects the current lack of theoretical understanding of TFBS evolution in the literature, even under the simplest case of directional selection.
Basic population genetics shows that directional selection is expected to cause a change, e.g., yield a functional binding site, over times on the order of $1/(NsU_{b})$, where $N$ is the population size, $s$ is the selection advantage of a binding site, and $U_{b}$ is the beneficial mutation rate~\cite{berg_adaptive_2004}.
This process can be extremely slow, especially under neutrality, if several mutational steps are needed to reach a sequence with sufficient binding energy to confer a selective advantage. 
As already pointed out by Berg \textit{et al.} (2004)~\cite{berg_adaptive_2004}, this places strong constraints on the length of the binding sites, if they were to evolve from random sequences. 

Several biophysical factors, such as TF concentration and the energetics of TF-DNA and TF-TF interactions, might play an important role in TFBS evolution. 
Quantitative models for TF sequence specificity~\cite{von_hippel_specificity_1986, berg_selection_1987, stormo_specificity_1998, stormo_identifying_1989, stormo_determining_2010, zhao_inferring_2009} and for thermodynamic (TD) equilibrium of TF occupancy on DNA~\cite{shea_or_1984, berg_selection_1987, bintu_transcriptional_2005-1, bintu_transcriptional_2005, hermsen_transcriptional_2006, hermsen_combinatorial_2010} were developed in recent decades and, in parallel with developments in sequencing, have contributed to our understanding of TF-DNA interaction biophysics.
These biophysical factors can shape the characteristics of the TFBS fitness landscape over genotype space in evolutionary models~\cite{gerland_physical_2002,gerland_selection_2002, macarthur_expected_2004, berg_adaptive_2004,stewart_why_2012,stewart_evolution_2013,payne_robustness_2014}. There are also intensive efforts to understand the mapping from promoter/enhancer sequences to gene expression~\cite{segal_predicting_2008, hermsen_transcriptional_2006, samee_quantitative_2014, he_thermodynamics-based_2010}.
Despite this recent attention, there have been relatively few attempts to understand the evolutionary dynamics of TFBS in full promoter/enhancer regions~\cite{macarthur_expected_2004, he_evolutionary_2012, hermsen_combinatorial_2010, duque_simulations_2013, duque_what_2015}, especially using biophysically realistic but still mathematically tractable models.  Such models are necessary to gain a thorough theoretical understanding of binding site evolution.

Our aim in this study is to investigate the dynamics of TFBS evolution by focusing on the typical evolutionary rates for individual TFBS gain and loss events. 
We consider both a single binding site at an isolated DNA region and a full enhancer/promoter region, able to harbor multiple binding sites. 
In the following section, we lay out our modeling framework, which covers both population genetic and biophysical considerations, as outlined above. 
Using this framework, we try to understand 
{\bf i)} what typical gain and loss rates are for a single TFBS site; 
{\bf ii)} how quickly populations converge to a stationary distribution for a single TFBS; 
{\bf iii)} how multiple TFBS evolve in enhancers and promoters; 
{\bf iv)} how early history of the evolving sequences can change the evolutionary rates of TFBS; and 
{\bf v)} how cooperativity between TFs affects the evolution of gene expression. 
We find that, under realistic parameter ranges, both gain and loss of a single binding site is slow, slower than the typical divergence time between species.
Importantly, fast emergence of an isolated TFBS requires strong selection and favorable initial sequences in the mutational neighborhood of a strong TFBS.
The evolutionary process approaches the equilibrium distribution very slowly, raising concerns about the use of equilibrium assumptions in theoretical work. We proceed to show that the dynamics of TFBS
evolution in larger sequences can be understood approximately from the dynamics of single binding sites; the TFBS gain times are again slow if evolution starts from random sequence in the absence of strong selection or large regulatory sequence ``real estate.''
Finally, we identify two factors that can speed up the emergence of TFBS: the existence of an initial sequence distribution biased towards the mutational neighborhood of strongly binding sequences, which suggests that ancient evolutionary history can play a major role in the emergence of ``novelties''~\cite{villar_enhancer_2015}; and the biophysical cooperativity between transcription factors, which can partially account for the lack of observed correlation between identifiable binding sequences and transcriptional activity~\cite{stefflova_cooperativity_2013}.
\section{Models \& Methods}

\subsection{Population genetics}

\begin{figure}
\begin{centering}
\includegraphics[height=0.75\textheight]{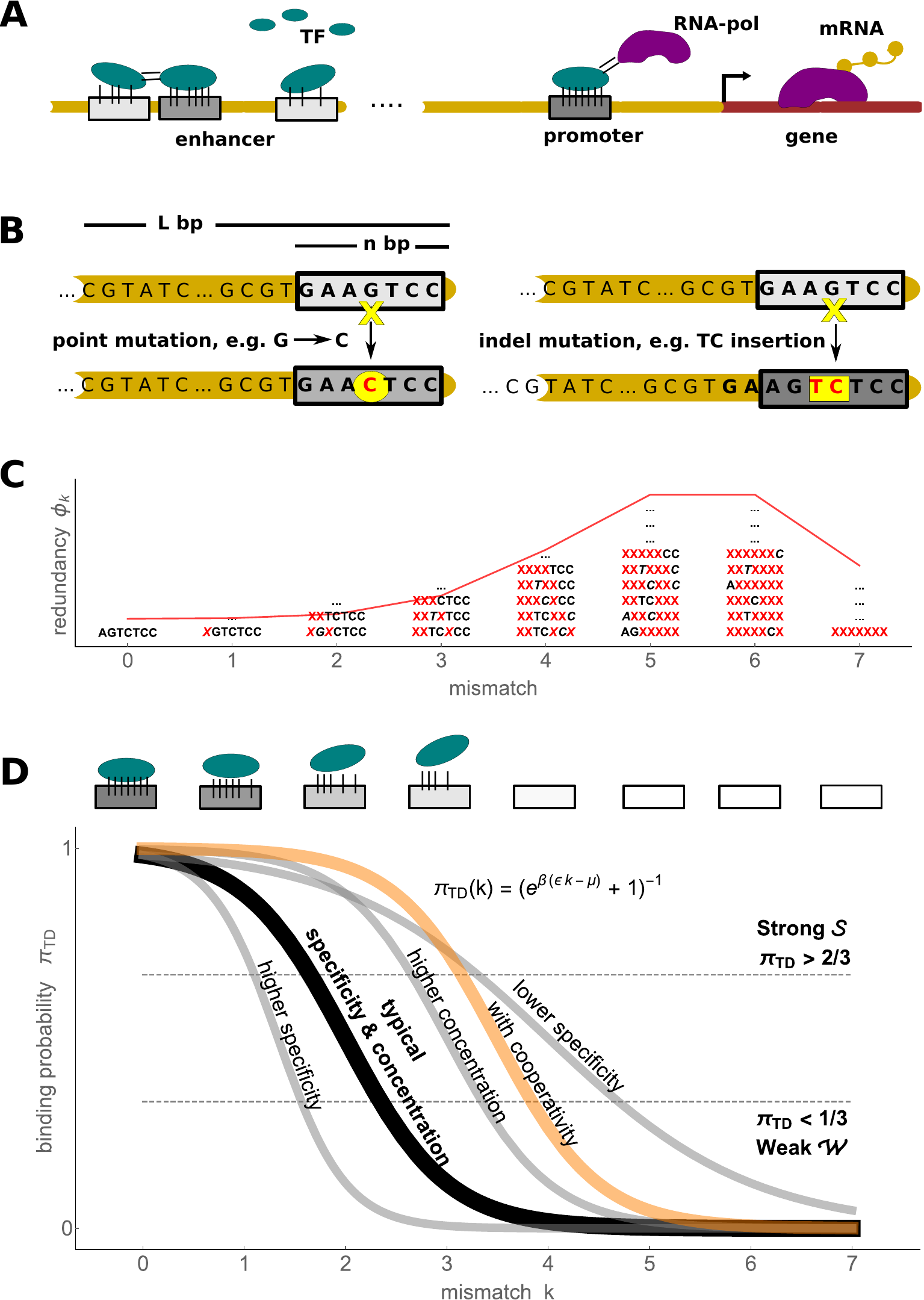}
\end{centering}
\caption{
\textbf{Biophysics of transcription regulation.}
\textbf{A)}
TFs bind to regulatory DNA regions (promoters and enhancers) in a sequence-specific manner to regulate transcriptional gene expression
(mRNA production) level via different mechanisms, such as recruiting RNA polymerase (RNA-pol).
\textbf{B)}
A schematic of two types of mutational processes that we model: point mutations (left) and indel mutations (right). 
\textbf{C)}
The mismatch binding model results in redundancy of genotype classes, with a binomial distribution (red) of genotypes in each mismatch class (some examples of degenerate sequences shown)
\textbf{D)}
The mapping from the TFBS regulatory sequence to gene expression level is determined by the thermodynamic occupancy (binding
probability) of the binding site. 
If each of the $k$ mismatches from the consensus sequence decreases the binding energy by $\epsilon$, the occupancy of the binding site is
$\pi_{\rm TD}(k) = (1+e^{\beta(\epsilon k-\mu)})^{-1}$, where $\mu$ is the chemical potential (related to free TF concentration). A typical occupancy curve is shown in black ($\epsilon=2\,k_BT$ and $\mu=4\,k_BT$); the gray curves show the effect of perturbation to these parameters ($\epsilon=1\,k_BT$, $\epsilon=3\,k_BT$ and $\mu=6\,k_BT$); the orange curve illustrates the case of two cooperatively binding TFs ($k_c=0$ and $E_c=-3\,k_BT$, see text for details).
We pick two thresholds, shown in dashed lines, to define discrete binding classes: strong $\mathcal{S}$ ($\pi_{\rm TD}>2/3$) and weak $\mathcal{W}$ ($\pi_{\rm TD}<1/3$).
\label{Fig1}}
\end{figure}

We consider a finite population of $N$ diploid individuals whose genetic
content consists of an evolvable $L$ base pair (bp) contiguous regulatory
sequence $\boldsymbol \sigma$ to which TFs can bind. Given that
$\sigma_{i}\in\{A,\, C,\, G,\, T\}$ where $i =1,\, 2, \, ..., \, L$ indexes the position
in regulatory sequence, there are $4^{L}$
different regulatory sequences in the genotype space. Each TF is assumed
to bind to a contiguous sequence of $n$ bp within our focal region
of $L$ bp (Fig.~\ref{Fig1}A,B). Regulatory sequences evolve under mutation, selection,
and sampling drift. The rest of the genome is assumed to be identical for all individuals and is kept constant. In the first part of
our study we consider the regulatory sequence comprised of a single
TFBS (i.e. $L=n$). Later, we consider the evolution of a longer
sequence (i.e. $L \gg n$) in which more than one TFBS can evolve.
For simulations, we use a Wright-Fisher model where $N$ diploid individuals
are sampled from the previous generation after mutation and selection.
Our analytical treatment is general and corresponds to setups where a diffusion approximation to allele frequency
evolution is valid. We neglect recombination since typical regulatory
sequences are short, $L\leq1000$. To be consistent with most of the population genetics literature we assume diploidy, but since we do not consider any dominance effects, our  results 
also hold for a haploid population with $2N$ individuals.

Evolutionary dynamics simplify
in the low mutation limit where the
population consists of a single genotype during most of its evolutionary
history (the fixed state population model). Desai \& Fisher~\cite{desai_beneficial_2007}
have shown that the condition $\frac{\log 4N \Delta f}{\Delta f} \ll \frac{1}{4NU_b \Delta f}$
needs to hold for a fixed state population assumption to be accurate. The term on the left is
the establishment time of a mutant allele with a selective advantage
$\Delta f$ relative to the wild type;  the term on the right-hand side is the waiting time
for such an allele to appear, where $U_b$ is the beneficial mutation
rate per individual per generation. Note that, in binding site context, $U_b$ refers to the rate of mutations which increase the fitness, for instance, by increasing binding strength. Its exact value depends on the current state of the genotype; nevertheless, typical value estimates help model the evolutionary dynamics. In multicellular eukaryotes, where most evidence for the evolution of TFBSs has been collected and which provide the motivation for this manuscript, the number of mutations per nucleotide site is
typically low, e.g. $4Nu \sim 0.01$ in \emph{Drosophila} and $4Nu \sim 0.001$
in humans~\cite{lynch_origins_2003}, where $u$ is the point mutation rate per generation per base pair. For a single binding
site of typical length $n \sim 5-15$, one therefore expects the fixed
state population model to be accurate. For longer regulatory sequences, one expects
that beneficial mutations are rare among all possible mutations, so that the fixed state population model can be assumed to hold as well.

Evolution under the fixed state assumption can be treated
as a simple Markovian jump process. The transition rate from a regulatory
sequence $\boldsymbol \sigma$ to another regulatory sequence $\boldsymbol \sigma ' $
in a diploid population is 
\begin{equation}
R_{ \sigma ', \sigma }=2N\; U_{ \sigma ', \sigma }\; P_{\rm fix}(N,\, \Delta f_{\sigma', \sigma})
\end{equation}
where $\Delta f_{\sigma ', \sigma} = f(\boldsymbol \sigma' ) - f(\boldsymbol \sigma)$
is the fitness difference and $U_{\sigma', \sigma}$
is the mutation rate from $\boldsymbol \sigma$ to $\boldsymbol \sigma'$. The fixation probability $P_{\rm fix}$ of a mutation with fitness difference $\Delta f$ in a diploid population of $N$ individuals is 
\begin{equation}
P_{\rm fix}(N,\, \Delta f) = \frac{1-e^{-2\Delta f}}{1-e^{-4N\Delta f}} \approx \frac{2\Delta f}{1-e^{-4N\Delta f}},
\end{equation}
which is based on the diffusion approximation~\cite{kimura_probability_1962}. Note that the fixation probability scaled with $1/N$ approximates to $2N\Delta f$ when  $N\Delta f \gg 1$.
Evolutionary dynamics therefore depend essentially on how regulatory
sequences are mutationally connected in genotype space, and how fitnesses
differ between neighboring genotypes, i.e., on the fitness landscape.

\subsection{Directional selection on biophysically motivated fitness landscapes}

In this study, we focus on directional selection by assuming that fitness $f$ is proportional to gene expression level $g$ which depends on regulatory sequence, i.e.
\begin{equation}
f(\boldsymbol \sigma )=s\,g(\boldsymbol \sigma)
\label{eq:DirectionalSelection}
\end{equation}
where $s$ is the selection strength. It is important to note that this choice does not imply that directional selection is the only natural selection mechanism. 
It simply aims at obtaining the theoretical upper limits for the rates of gaining and losing binding sites.

To analyze a realistic but tractable mapping from the regulatory
sequence to fitness, we primarily assume that the proxy for gene expression is the binding occupancy (binding probability) $\pi$ at a single TF binding site, or the sum of the binding occupancies  within an enhancer/promoter region (based on limited experimental support~\cite{giorgetti_noncooperative_2010}). This corresponds to
\begin{equation}
f(\boldsymbol \sigma) = s \sum_i \pi^{(i)}(\boldsymbol \sigma)
\label{GeneExpressionModel1}
\end{equation}
where $\pi^{(i)}$ is the binding occupancy of a site starting at the nucleotide $i$ in sequence $\boldsymbol \sigma$, and $s$ can be interpreted as the selective advantage of a strongest binding to a weakest binding at a site. We assume all binding sites have equal strength and direction in their contribution towards total gene activation.  Sites acting as repressors in our simple model would enter into Eq~(\ref{GeneExpressionModel1}) with a negative selection strength, $s$. Future studies developing mathematically tractable models should consider more realistic case of unequal contribution with combined activator and repressor sites responding differentially to various regulatory inputs~\cite{duque_what_2015}.  Although one can postulate different scenarios that map TF occupancies in a long ($L\gg n$) promoter to gene expression, we chose the simplest case which allows us to make analytical calculations. Later we relax our assumption on noninteracting binding sites and consider the effects of several kinds of interactions on gene expression and thus on evolutionary dynamics.

The occupancy of the TF on its binding site is assumed to be in thermodynamic (TD) equilibrium~\cite{shea_or_1984, berg_selection_1987, bintu_transcriptional_2005-1, bintu_transcriptional_2005, hermsen_transcriptional_2006, hermsen_combinatorial_2010}.
While this might not always be realistic \cite{hammar_direct_2014, cepeda-humerez_stochastic_2015}, there is empirical support for this assumption (particularly in prokaryotes) \cite{segal_predicting_2008, brewster_tuning_2012, razo-mejia_comparison_2014},
and more importantly, it is sufficient to capture the essential nonlinearity in this genotype-phenotype-fitness mapping \cite{haldane_biophysical_2014}.
In thermodynamic equilibrium, the binding occupancy at the site starting with the $i$-th position in regulatory sequence is given by
\begin{equation}
\pi_{\rm TD}^{(i)}(E_i) = \Big(1 + e^{\beta(E_i - \mu)} \Big)^{-1}.
\label{eq:TDmodel}
\end{equation}
Here, $\mu$ is the chemical potential of the TF (related to its free concentration) \cite{gerland_physical_2002,weinert_scaling_2014}; $E_{i}$ is the sequence specific binding energy, where lower energy corresponds to tighter binding, and $\beta=(k_B T)^{-1}$. 
We compute the binding energy $E_i$ by adopting an additive energy model 
which is considered to be valid at least up to a few mismatches from the consensus
sequence \cite{maerkl_systems_2007, zhao_inferring_2009, stormo_determining_2010, kinney_using_2010}, i.e.
\begin{equation}
E_i(\boldsymbol \sigma ) = \sum_{j=i}^{i+n-1} \xi_{\sigma_j, j}
\end{equation}
where $\xi$ stands for the energy matrix whose $\xi_{\sigma_j, j}$
element gives the energetic contribution of the nucleotide $\sigma_j$
appearing at the $j$-th position within TFBS.
With this, Eq~(\ref{GeneExpressionModel1}) can be rewritten more formally as
\begin{equation}
f(\boldsymbol \sigma) = s \sum_i \pi_{\rm TD}^{(i)} (E_i(\boldsymbol \sigma))
\label{GeneExpressionModel}
\end{equation}

To allow analytical progress, we make the ``mismatch assumption,'' i.e., the energy matrices contain identical $\epsilon >0$ entries for every non-consensus (mismatch) base pair; the consensus entries are set to zero by convention. A 
single binding sequence with $k$ mismatches therefore has the binding energy $E = k \epsilon$. We will refer to $\epsilon$ as ``specificity.''
Specificity is provided by diverse interactions between DNA and TF, including specific hydrogen bonds, van der Waals forces, steric exclusions, unpaired polar atoms, etc.~\cite{mckeown_evolution_2014}.
$\epsilon$ is expected to be in the range $1-3$ $k_{B}T$, which is consistent with theoretical arguments \cite{gerland_physical_2002} as well as direct measurements \cite{fields_quantitative_1997, kinney_using_2010, maerkl_systems_2007}. Note that we explicitly check the validity of the analytical results based on the mismatch assumption by comparing them against simulations using realistic energy matrices.
The redundancy (i.e., normalized number of distinct sequences) of a mismatch class $k$ at a
single site in a random genome can be described by a binomial distribution
$\boldsymbol \phi$ (Fig.~\ref{Fig1}C) where the probability of encountering a mismatch class $k$ is
\begin{equation}
\boldsymbol{\phi_{k}}(n,\alpha) = \binom{n}{k}\; \alpha^{k} \big( 1 - \alpha \big)^{n-k}
\label{eq:binomialdistr}
\end{equation}
where $\alpha=3/4$ in the case of equiprobable distribution over the four nucleotides.

We focus on selection in a single environment, which in this framework corresponds to a single choice for the TF concentration.
We therefore fix the chemical potential to a baseline value of $\mu=4\, k_{B}T$, which maps changes in the sequence (mismatch class $k$) to a full range of gene expression levels, as shown in Fig.~\ref{Fig1}D. We subsequently vary $\mu$ systematically and report how its value affects the results.

After these preliminaries, the equilibrium binding probability of Eq~(\ref{eq:TDmodel}) reduces to
\begin{equation}
\pi_{{\rm TD}}(k) = \Big( 1 + e^{\beta(\epsilon\; k - \mu)} \Big)^{-1}.
\end{equation}
This function has a sigmoid shape whose steepness depends on specificity
$\epsilon$ and whose midpoint depends on the ratio of chemical potential
to specificity, $\mu / \epsilon$ (Fig.~\ref{Fig1}D).
To simplify discussion, we introduce two classes of sequences: 
genotypes are associated with ``strong binding'' $\mathcal S$
and ``weak binding'' $\mathcal W$ if $\pi_{\rm TD} > 2/3$ and $\pi_{\rm TD} < 1/3$,
respectively. The thresholds that we pick are arbitrary, while still placing the midpoint of the sigmoid between the two classes; our results do not change qualitatively for
other choices of thresholds. In the mismatch approximation, the genotype
classes $k= \{ 0,\,1,\,...,\, k_{\mathcal S} \} \in \mathcal S$
and $k= \{ k_{\mathcal W},\, k_{\mathcal W} + 1, \, ..., \, n \} \in \mathcal W$
correspond to strong and weak binding, respectively.  $k_{\mathcal{S}}$
and $k_{\mathcal W}$ are defined as the closest integers to the thresholds defined above; these values depend on $\epsilon$ and $\mu$. 
We also define a ``presite'' as the mismatch class that is $1$ bp away from the threshold for strong binding, i.e., a class with $k_{\mathcal S} + 1$ mismatches.
Note that binding length $n$ extends the tail of the fitness landscape
for a single site and shifts the center of redundancy rich mismatch
classes  (Fig.~\ref{Fig1}C).

The formulation in Eq~(\ref{GeneExpressionModel}) reduces to 
\begin{equation}
f(k) = s\,\pi_{\rm TD}(k)
\end{equation}
in a mismatch approximation at a single site, which we will investigate extensively for $Ns$ scaling of TFBS gain and loss rates.
We consider a wide range of $Ns$ values: $Ns<0$ for negative selection, $Ns=0$ for neutral evolution, $Ns\sim1$ for weak positive selection, $Ns \gg n \log(2)/2$ for strong positive selection (see below for this particular choice of the threshold). 

In order to study the effects of interacting TFBSs in large regulatory sequences, we relax our assumption of non-interacting TFBS in Eq~(\ref{GeneExpressionModel}) and study three simple models.
In the main text, we report the cooperative physical interaction between two TF molecules binding two nearby sites where the binding probability at a site is modified as
\begin{equation}
\pi_{\rm coop}(k,\,k_c) = \frac{e^{-\beta(\epsilon k - \mu)} + e^{-\beta (\epsilon ( k + k_c ) - 2 \mu - E_c)}}{1 + e^{-\beta ( \epsilon k - \mu)} + e^{-\beta (\epsilon k_c - \mu)} + e^{-\beta (\epsilon ( k + k_c ) - 2 \mu - E_c)}},
\label{eq:TDmodelwcoop}
\end{equation}
where $k_c$  stands for the mismatch class at the co-binding
site and $E_c$ for cooperativity. In this study we consider that cooperative energy ranges from an intermediate strength ($E_{c}=-2\,k_BT$) to a high strength ($E_{c}=-4\,k_BT$)~\cite{hermsen_transcriptional_2006}.
Fig~\ref{Fig1}D shows an example of the binding probability when a strong co-binding site exists. 
As a function of $k$ alone, at fixed $k_c$, this formulation of cooperativity is consistent with the zero-cooperativity ($E_c=0$) case but with a changed effective chemical potential.
We take cooperative interactions into account if the two TFs are binding within $3$ bp of each other, and we only consider the
strongest binding of the cooperative partner (i.e., the proximal location with the lowest $k_c$).

In Supporting Information, we discuss the other two models of interacting TFBS. In one model, gene expression is determined only by the binding probability of the strongest site in the regulatory sequence. In the other model, gene expression is determined by the probability of the joint occupancy of $2$ strongest binding sites, anywhere in the regulatory sequence; this model is a toy version of synergistic ``non-physical'' interaction of TFs which compete with nucleosomal binding for the occupancy of regulatory regions in eukaryotes (see Mirny (2010)~\cite{mirny_nucleosome-mediated_2010} for a detailed model).

\subsection{Point and indel mutations}

Point mutations and indels are the only mutational processes in our framework. Point mutations with a rate $u$ convert the nucleotide at
one position into one of the $3$ other nucleotide types. For a single
binding site, the probability that a point mutation changes the mismatch
class from $k$ to $k'$ is 

\begin{equation}
\boldsymbol{P}_{k',k}^{\rm (point)}=\big(1-k/n\big)\,\delta_{k',k+1}+\big(k/3n\big)\,\delta_{k',k-1}+\big(2k/3n\big)\,\delta_{k',k}
\end{equation}
where $\delta_{a,b}=1$ if
$a=b$ and $0$ otherwise.

We define the indel mutation rate per base pair such that it occurs with rate $\theta\, u$ at a position where a random nucleotide sequence is either inserted, or an existing nucleotide sequence is deleted. 
For mathematical simplicity, we assume that insertions and
deletions are equally likely; in fact, a slight bias towards deletions is reported in the literature with a ratio of deletion to insertion $\sim1.1-3.0$~\cite{taylor_occurrence_2004, brandstrom_genomic_2007, park_ancestral_2015}. 
Parameter $\theta$ is the ratio of indel mutation rate to point mutation rate, and is reported to be in the range $0.1-0.2$~\cite{cartwright_problems_2009, chen_variation_2009, lee_rate_2012}. 
We consider two cases: the baseline of $\theta=0$ for
no indel mutations, and $\theta=0.15$ for the combined effect of indel and point mutations. 
Since we fix the length of the regulatory sequence, indels shift existing positions 
away from or inwards to some reference position (e.g., transcription start site). For consistency,
we fix the regulatory sequence at its final position and assume
that sequences before the initial position are random. 
Indel lengths vary, with reports suggesting a sharply decreasing but fat-tail frequency distribution~\cite{keightley_mcalign:_2004}. For simulations 
we consider only very short indels of size $1-2$ bp, occurring proportional with their reported frequencies of $0.45$ and $0.18$, respectively. We do not need to assume any particular indel length for analytical calculations (below). While sufficient for our purposes, this setup would need to be modified when working with real sequence alignments of orthologous regions.

For a single binding site (i.e. $L=n$) one can exactly calculate the probability of an indel mutation changing
the mismatch class from $k$ to $k'$ as
\begin{equation}
\boldsymbol{P}_{k',k}^{\rm (indel)} = \sum_{i=1}^{n}(1/n)\; \sum_{x=0}^{k'} \; p(X_i = x \; | \; k)\; p(Y_i = k' - x).\label{eq:indel-mutation}
\end{equation}
Here, $i$ is the index for the position of an indel mutation within the
binding site. The distribution over possible positions is uniform (hence $1/n$). The indel mutation defines two distinct
parts in the binding site in terms of mismatches: nucleotides behind
the indel mutation preserve their mismatch information, yet the nucleotides
within and after indel mutation completely lose it. The new mismatches at these distinct parts $X_i$ and $Y_i$
are binomial random variables,
\begin{equation}
\begin{array}{lcl}
p(X_i = x\; | \; k) & = &\boldsymbol{\phi_x} (i-1, \, \alpha = k/n) \\
p(Y_i = y) & = & \boldsymbol{\phi_y} (n-i+1,\, \alpha = 3/4)
\end{array}
\end{equation}
where $\boldsymbol{\phi_k}(n,\,\alpha)$ is defined in Eq~(\ref{eq:binomialdistr}). Fig~\ref{FigS1} shows that Monte Carlo sampling of indel mutations at a single binding site
matches the analytical expression in Eq~(\ref{eq:indel-mutation}). 

The two types of mutations can be combined into the mutation rate matrix as follows:
\begin{equation}
\boldsymbol{U}_{k',k} = 
\begin{cases}
n\; u\; \Big( \boldsymbol P_{k', k}^{\rm (point)} + \theta \; \boldsymbol P_{k',k}^{\rm (indel)} \Big)   &   k'  \neq k  \\
-\sum_{ k' \neq k} \boldsymbol U_{k', k}   &   k'=k
\end{cases}.
\end{equation}

\subsection{Evolutionary dynamics of single TF binding sites}

For a sequence that consists of an isolated TFBS (i.e., $L=n$),  analytical treatment is possible under the fixed  state  assumption. 
Let $\boldsymbol \psi (t)$ be a distribution over an ensemble of populations, whose $k$-th component, $\boldsymbol \psi_k(t)$,
denotes the probability of detecting a genotype with $k$ mismatches
at time $t$. In the continuous time limit, the evolution of $\boldsymbol \psi (t)$ is described by
\begin{equation}
\frac{d}{dt} \boldsymbol \psi(t) = \boldsymbol R \cdot \boldsymbol \psi
\end{equation}
which accepts the following solution:
\begin{equation}
\boldsymbol \psi (t) = e^{\boldsymbol R \; t} \cdot \boldsymbol \psi (0).
\label{eq:timesolutionR}
\end{equation}
Here, $\boldsymbol R $ is the transition rate matrix defined as
\begin{equation}
\boldsymbol R_{k',k} =
\begin{cases}
2N \; \boldsymbol U_{k',k} \; P_{\rm fix} (N,\, \Delta f_{k',k}) &  k' \neq k \\
- \sum_{k' \neq k} \boldsymbol R_{k',k}  &  k' = k
\end{cases}.
\label{transitionr}
\end{equation}

This dynamical system is a continous-time Markov chain and there exists a unique stationary distribution $\hat{\boldsymbol{\psi}}$ corresponding the genotype distribution over an ensemble of populations at large time points. It can be calculated by decomposing the transition rate matrix $\boldsymbol R $ into its eigenvalues and eigenvectors. The normalised left eigenvector with zero eigenvalue corresponds to the stationary distribution. This can also be expressed
analytically as
\begin{equation}
\hat{\boldsymbol \psi}_k \propto e^{F(k,N) + H(k\; | \; n, \alpha)},
\label{equilibriumsoln}
\end{equation}
where $F(k,N) = 4 N f(k)$ captures the relative importance of 
selection to genetic drift, and $H(k\; | \; n,\alpha)$ is the mutational
entropy, describing how a particular mismatch class $k$ is favored
due to redundancy and connectivity of the genotype space. For point mutations alone ($\theta=0$), $H=\log\phi_{k}(n,\alpha)$, with the binomial distribution $\phi_{k}(n,\; \alpha)$ as defined in Eq~(\ref{eq:binomialdistr}). Obtaining a closed form expression for $H$ is difficult when considering indel mutations ($\theta>0$), yet the  eigenvalue
method solution suggests a similar shape for $\theta$ in the range of interest. 
The form of the stationary distribution was known for a long time in population genetics literature for a single locus or many loci with linkage equilibrium~\cite{wright_evolution_1931}. It has recently been generalised to arbitrary sequence space under the fixed state assumption~\cite{berg_adaptive_2004, sella_application_2005}, resulting in the form of Eq~(\ref{equilibriumsoln}) with a close analogy in the energy-entropy balance of statistical physics~\cite{barton_application_2009}, and become a subject of theoretical interest~\cite{mustonen_evolutionary_2005, mustonen_energy-dependent_2008, manhart_universal_2012, haldane_biophysical_2014}. 

Under weak directional selection for high expression (and thus high binding site occupancy),
the stationary distribution shows a bimodal shape, with one peak located around the fittest class, $k \sim 0$, and another
at the core of mutational entropy, $k \sim \alpha \, n$ (recall that
$\alpha=3/4$ for a completely random genome). This
bimodal shape collapses to a unimodal one, either at no selection or at strong selection. The threshold value for $Ns$
distinguishing strong and weak selection regimes primarily depends on the
TFBS binding length, $n$. In a sigmoidal fitness landscape and approximating the binomial distribution by a normal
distribution as appropriate, the sizes of these two peaks are roughly
proportional to $\exp \left(4Ns-n\log4\right)$ and $\sqrt{2\pi\alpha(1-\alpha)n}$,
respectively. Therefore, we expect  the threshold $Ns$ to scale
as $\frac{1}{4}\Big(n\,\log4-\frac{1}{2}\log2\pi\alpha(1-\alpha)n\Big)$.
For typical $n$, the linear term is dominant, suggesting that

\begin{equation}
Ns \sim n \log (2)/2
\label{eq:thresholdNS}
\end{equation}
corresponds to the threshold for strong selection in TFBS evolution (cf. Fig~\ref{FigS2}). Note that this $n$ scaling differs from the $\log(n)$  scaling which is expected in simple fitness landscapes~\cite{paixao_first_2015}.
Our argument assumes that the system is at evolutionary equilibrium, which, as we will see, is not necessarily the case even under strong selection, providing further motivation for focusing on dynamical aspects of evolution.

We define the time needed to gain (or lose) a TFBS as the time it takes for a strong binding site to emerge from a weak one (and vice versa), as schematized in Fig.~\ref{Fig1}D. 
For an isolated TFBS, these times can be computed from the Markovian properties of the evolutionary dynamics, by
calculating the average first hitting times~\cite{otto_biologists_2007}.
We will use the notations $\langle t \rangle_{\mathcal S \leftarrow k}$
and $\langle t \rangle_{\mathcal W \leftarrow k}$, respectively, for
average gain and loss times when evolution starts from mismatch class
$k$. Obviously, $\langle t \rangle_{\mathcal S \leftarrow k} = 0$ if
$k$ is among the strong binding classes ($k \in \mathcal S$)
and $\langle t \rangle_{\mathcal W \leftarrow k} = 0$ if $k$ is among
the weak binding classes ($k \in \mathcal W$). The average gain
times from other mismatch classes can be found by considering the relation $\langle t \rangle_{\mathcal S \leftarrow k} = 1+ \sum_{k' \notin \mathcal S} \boldsymbol P_{k,k'} \langle t \rangle_{\mathcal S \leftarrow k'}$, where $\boldsymbol P_{k,k'}$ is the probability of transition from $k'$ to $k$ in one generation. One can compute the average gain times by writing it in terms of linear algebraic equation:
\begin{equation}
\boldsymbol T_{\mathcal S \leftarrow} = \mathbf{(R_{\notin \mathcal S})}^{\rm -T} \cdot (-\boldsymbol 1)
\label{eq:averagetimematrix}
\end{equation}
where $ \boldsymbol T_{\mathcal S \leftarrow}$ is a column
vector listing non-trivial gain times, i.e. $ \{ \langle t \rangle_{S \leftarrow k} \}$
for $k = k_{\mathcal S} + 1, \; ..., \; n$. $\mathbf{R_{\notin \mathcal S}}$
is the $\mathbf R$ matrix with all rows and columns corresponding
to $k \in \mathcal S$  deleted and $-{\rm T}$ is the matrix operator
for the transpose after an inverse operation. $\boldsymbol 1$
is a vector of ones. Similarly one can find the loss times,
\begin{equation}
\boldsymbol T_{\mathcal W \leftarrow}  =  \mathbf{(R_{\notin \mathcal W })}^{\rm -T} \cdot (-\boldsymbol 1)
\label{eq:averagetimematrix-1}
\end{equation}
where $\boldsymbol T_{\mathcal W \leftarrow}$ is a column
vector listing non-trivial loss times, i.e. $\left\{ \langle t\rangle_{\mathcal{W}\leftarrow k}\right\} $
for $k=1,\; 2, \;...\; k_W - 1$. $\mathbf{R_{\notin \mathcal W}}$
is the $\mathbf R$ matrix with all rows and columns corresponding
to $k \in \mathcal W$  deleted.

In the case of point mutations alone ($\theta=0$), the
$\mathbf{R}$ matrix is tri-diagonal and one can deduce simpler
formulae for gain and loss times: 
\begin{equation}
\begin{array}{lcl}
\langle t \rangle_{\mathcal S \leftarrow k}^{\rm (point)} & = & \sum_{i=k_{\mathcal S} + 1}^k \frac{1}{\boldsymbol R_{i-1,\, i}} \; \frac{1 - \hat{\boldsymbol \Psi}_{i-1}}{\hat{\boldsymbol{\psi}}_{i}}\\
&&\\
\langle t \rangle_{\mathcal W \leftarrow k}^{\rm (point)} & =& \sum_{i=k+1}^{k_{\mathcal W}} \frac{1}{\boldsymbol R_{i-1,\, i}} \; \frac{\hat{\boldsymbol \Psi}_{i-1}}{\hat{\boldsymbol \psi}_i}
\label{eq:AverageTimeShortestPathGain-1}
\end{array}
\end{equation}
where we use $\hat{\boldsymbol \Psi}_i = \sum_{j=0}^i \; \hat{\boldsymbol \psi}_j$
to denote the cumulative stationary distribution. For very strong
selection, the second term in the sums approaches unity, resulting
in even simpler formulae~\cite{berg_adaptive_2004}, called the ``shortest path''
(sp) solution:
\begin{equation}
\begin{array}{lcl}
\langle t \rangle_{\mathcal S \leftarrow k}^{\rm (sp)} & = & \sum_{i = k_{\mathcal S} + 1}^k \; \frac{1}{\boldsymbol{R}_{i-1,\, i}} \\
&&\\
\langle t \rangle_{\mathcal W \leftarrow k}^{\rm (sp)} & = & \sum_{i = k+1}^{k_{\mathcal W}} \; \frac{1}{\boldsymbol{R}_{i-1,\, i}}
\end{array}.
\label{eq:AverageTimeShortestPathGain}
\end{equation}
These equations can be used to quickly estimate gain and loss rates of interest. For example, the gain rate from presites under strong selection is approximately $ 2  \,Ns\, u\frac{k_{\mathcal S}+1}{3} (f(k_{\mathcal S})-f(k_{\mathcal S}+1))$. Although the exact value depends on the binding specificity and chemical potential, one can see that it is about $Ns\,u$ for the parameter range of  interest. Similarly, one can see that the rate of loss from strong sites is about $2n\,|Ns|\,u$ when there is strong negative selection.

\section{Results}

\subsection{Single TF binding site gain and loss rates under mutation-selection-drift are typically slow}

We first studied the evolutionary rates for a single TF binding site at an isolated DNA sequence of the same length under mutation, genetic drift, and directional selection for high gene expression level (i.e., tighter binding).
As detailed in the Models \& Methods section, we combined a thermodynamically motivated fitness landscape with the mismatch approximation, and assumed that the mutation rate is low enough for the fixed state population approximation to be valid. Under these assumptions, we could calculate the inverse of the average TFBS gain and loss times as a function of the starting genotype, using either an exact method or Wright-Fisher simulations. We considered point mutations alone, or point mutations combined with short indel mutations, in order to understand under which conditions the rates of gaining and losing binding sites can reach or exceed the rates $2-3$ orders of magnitude greater than point mutation rate, and thus to become  comparable to rates observed in comparative genomic studies.

\begin{figure}
\begin{centering}
\includegraphics[width=\textwidth]{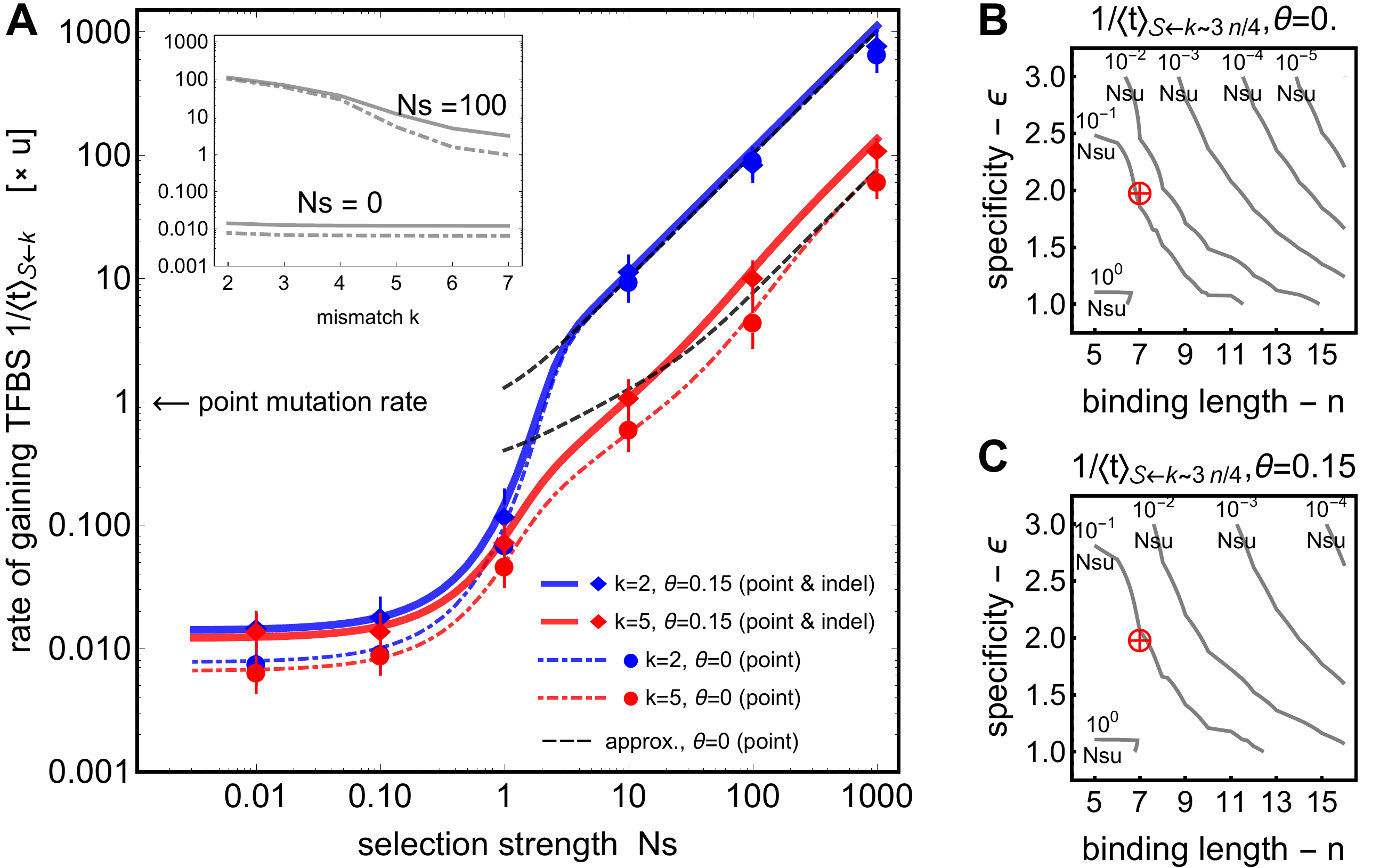}
\end{centering}
\caption{
\textbf{Single TF binding site gain rates at an isolated DNA region. A)}
The dependence of the gain rate, $1/\langle t\rangle_{\mathcal S \leftarrow k}$ shown in units of point mutation rate, from sequences in different initial mismatch classes $k$ (blue: $k=2$, red: $k=5$),
as a function of selection strength. Results with point mutations only ($\theta=0$) are shown by dashed line; with admixture of indel mutations ($\theta=0.15$) by a solid line. For strong selection, $Ns\gg n\log(2)/2$, the rates scale with $Ns$, which is captured well by the ``shortest path'' approximation (black dashed lines in the main figure) of Eq~(\ref{eq:AverageTimeShortestPathGain}). The biophysical parameters are:
site length $n=7$ bp; binding specificity $\epsilon=2$ $k_B T$; chemical
potential $\mu=4$ $k_{B}T$. Points correspond to Wright-Fisher simulations with $Nu=0.01$ where error bars cover $\pm 2$ SEM (standard error of mean).
Inset shows the behavior of the gain rates as a function of the initial mismatch class $k$ for $Ns=0$
and $Ns=100$.
\textbf{B, C)} Gain rates from redundancy rich classes ($k\sim3n/4$, typical of evolution from random ``virgin'' sequence) under strong selection, without (B) and with (C) indel mutations supplementing the point mutations. Red crosshairs denote the cases depicted in panel A. Contour lines show constant gain rates in units of $Ns\,u$ as a function of biophysical parameters $n$ and $\epsilon$. Wiggles in the contour lines are not a numerical artefact but a consequence of discrete mismatch classes. 
\label{Fig2}}
\end{figure}

Fig.~\ref{Fig2}A shows the dependence of the TFBS gain rate on the selection strength (with respect to genetic drift), $Ns$. For parameters typical of eukaryotic binding sites (length $n=7$ bp, specificity $\epsilon=2\,k_BT$), the TFBS gain rates are extremely slow (practically no evolution) when there is negligible selection pressure ($Ns\sim 0$), indicating the importance of selection for TFBS emergence. Indeed, the effective selection needs to be very strong, e.g., $Ns>100$, for TFBS evolution to exceed the per-nucleotide mutation rate by orders of magnitude and become comparable to speciation rates. 

Even if strong selection were present, the gain rate depends crucially on the initial genotype. While gain rates from presites, i.e., genotypes one mutation away from the threshold for strong binding, are roughly $Ns\,u$ for the strong $Ns$ regime (as estimated by Berg \textit{et al.}~\cite{berg_adaptive_2004}), they decrease dramatically if more mutational steps are needed to evolve a functionally strong binding site. This is illustrated in the inset to Fig~\ref{Fig2}A, showing an exponential-like decay in the gain rates as a function of the number of mismatches, even for a TFBS of a modest length of 7 bp. As argued in the Models \& Methods section (see Eq~(\ref{eq:thresholdNS})), we confirmed that the threshold for the strong $Ns$ regime scales as $n\log(2)/2$ and not as $\log(n)$ which is the case for simple fitness landscapes~\cite{paixao_first_2015}. 

The availability of a realistic fraction of indel mutations (here, $\theta=0.15$) can speed up  evolution when starting from distant genotypes (cf. solid and dashed red line in Fig~\ref{Fig2}A). This is  because indels connect the genotype space such that paths from many to few mismatches are possible within a single mutational step. Nevertheless, the improvement due to indel mutations does not alleviate the need for very strong selective pressure and the proximity of the initial to strongly-binding sequence, in order to evolve a functional site.

Biophysical parameters---the binding site length $n$, the chemical potential $\mu$, and the specificity $\epsilon$---influence the shape of the fitness landscape and thus the TFBS gain rates. This is especially evident when we consider \emph{de novo} evolution starting from random sequence.  As shown in Figs~\ref{Fig2}B,\,C, increases in specificity or length cause a sharp drop in the gain rates from initial sequences in the most redundancy rich class, which can be only partially mitigated by the availability of indel mutations. This especially suggests that adaptation of TFBS from random sequences for TF with very large binding lengths and very strong specificities is unlikely with point and indel mutations which can constrain the evolution of TF lengths and TF specificity, which is consistent with Berg {\it et al.} (2004)~\cite{berg_adaptive_2004}'s earlier numeric observation. Importantly, the binding specificity and length show an inverse relation with the logarithm of the gain rates. This is due to the fact that a decrease in specificity allows more genotypes to generate appreciable binding and therefore fitness (see Fig.~\ref{Fig1}D), which partially compensates the increase in mutational entropy at larger binding site lengths. Variation of the chemical potential $\mu$ corresponding to an order-of-magnitude change in the free TF concentration does not qualitatively affect the results.

Typically slow TFBS evolution is a consequence of the sigmoidal shape of the thermodynamically motivated fitness landscape, where adaptive evolution in the redundant but weakly binding classes $\mathcal{W}$ must proceed very slowly due to the absence of a selection gradient. To illustrate this point, we generated alternative fitness landscapes that agree exactly with the thermodynamically motivated one from the fittest class to the threshold class for strong binding, $k_{\mathcal S}$, but after that decay as power laws, $\pi_{\rm pl}$, with a tunable exponent (see SI text). As seen in Fig~\ref{FigS3}, this exponent is a major determinant of the gain rates, suggesting that a biophysically realistic fitness landscape is crucial for the quantitative understanding of TFBS evolution. 

To check that the assumption of the fixed state population is valid at $Nu=0.01$, the value used here that is also relevant for multicellular eukaryotes \cite{lynch_origins_2003}, we performed Wright-Fisher  simulations as described in the Models \& Methods section. Fig~\ref{Fig2}A shows excellent agreement between the analytical results and the simulation. We further increased the mutation rate to $Nu=0.1$, a regime more relevant for prokaryotes where polymorphisms in the population are no longer negligible, to find that the analytical fixed state assumption systematically overestimates the gain rates, as shown in Fig~\ref{FigS4}. In the presence of polymorphism, therefore, evolution at best proceeds as quickly as in monomorphic populations, and generally proceeds slower, so that our results provide a theoretical bound on the speed of adaptive evolution under directional selection. This is expected since the effects of clonal interference kick in after a certain $Nu$, where two different beneficial mutants start competing with each other, and eventually decrease the fixation probability in comparison to one beneficial mutant sweeping to fixation as in the monomorphic population case.

To check that the mismatch assumption does not strongly affect the reported results, we analyzed evolutionary dynamics with more realistic models of TF-DNA interaction. Different positions within the binding site can have different specificities, and one could suspect that this can significantly lower the evolutionary times. First, some positions within the TFBS may show almost no specificity for any nucleotide, most likely due to the geometry of TF-DNA interactions (e.g, when the TF can contact the nucleic acid residues only in the major groove); we have not simulated such cases explicitly, but simply take the binding site length $n$ to be the effective sequence length where TF does make specific contacts with the DNA. Second, the positions that do exhibit specificity might do so in a manner that is more inhomogeneous than our mismatch assumption, which assigns zero energy to the consensus and a constant $\epsilon$ to any possible mismatch. We thus generated energy matrices where $\epsilon$ was drawn from a Gaussian distribution with the same mean $\langle \epsilon \rangle=2\,k_BT$ as in our baseline case of Fig~\ref{Fig2}A, but with a standard deviation $0.5\,k_BT$. Fig~\ref{FigS5} shows that both equal and unequal energy contributions produce statistically similar behaviors, indicating that inhomogeneous binding interactions cannot substantially enhance the evolutionary rates.

We further investigated the rate of TFBS loss (Fig~\ref{FigS6}). Here too strong (negative) selection is needed to lose a site on reasonable timescales, and it is highly unlikely that a site would be lost in the presence of positive selection. In contrast to the TFBS gain case, however, negative selection and mutational entropy act in the same direction for TFBS loss, reducing the importance of the initial genotype and making selection more effective at larger $n$ and $\epsilon$.

Taken together, these results suggest that the emergence of an isolated TFBS under weak or no selection is typically slow relative to the species' divergence times, and gets rapidly slower for sites that are either longer or whose TFs are more specific than the baseline case considered here. This suggests that biophysical parameters themselves may be under evolutionary constraints; in particular, if point mutations and indels were the only mutational mechanisms, the evolution of long sites, e.g. $n\gg 10-12$, would seem extremely unlikely, as has been pointed out previously \cite{berg_adaptive_2004}. Absent any mechanisms that could lead to faster evolution and which we consider below, isolated TFBS are generally only likely to emerge in the presence of strong directional selection and a favorable distribution of initial sequences that is enriched in presites.

\subsection{Convergence to  the stationary distribution is slow and
depends strongly on initial conditions}

\begin{figure}
\begin{centering}
\includegraphics[width=\textwidth]{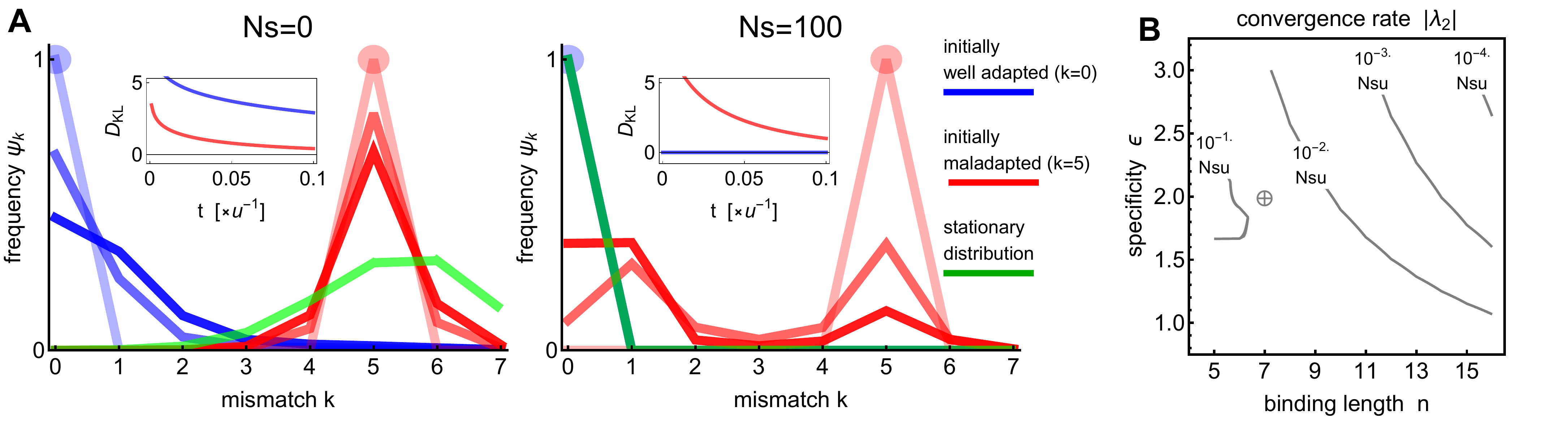}
\end{centering}
\caption{
\textbf{Convergence to the stationary distribution of TFBS sequences.}
\textbf{A)}
Evolutionary dynamics of the mismatch classes distribution $\boldsymbol{\psi}(k)$ for an isolated TFBS under point and indel mutations
($\theta=0.15$), directional selection for stronger binding,
and genetic drift is shown for initially well ($k=0$, blue) and badly
($k=5$, red) adapted populations. At left, no selection ($Ns=0$); at right, strong selection ($Ns=100$). Different curves show the distribution
of genotype classes at different time points ($t=0u^{-1},\; 0.05u^{-1},\; 0.1u^{-1}$ as decreasing opacity); stationary distribution is shown in green. Insets show the time evolution to convergence for initially well ($k=0$, blue) and badly ($k=5$,
red) adapted populations, measured by the Kullback-Leibler divergence $D_{KL}[\boldsymbol{\psi}(t)\,||\,\boldsymbol{\psi}(t=\infty)]$. The biophysical parameters are: $n=7$ bp, $\epsilon=2$ $k_BT$, $\mu=4$ $k_{B}T$. 
\textbf{B)}
Rate of convergence to the stationary distribution for different $\epsilon$ and $n$
values under strong selection ($Ns\gg n \log (2)/2$; here specifically $Ns=100$) and for $\theta=0.15$. Crosshairs represent the parameters used in a). 
\label{Fig3}}
\end{figure}

A number of previous studies (e.g., \cite{mustonen_evolutionary_2005, mustonen_energy-dependent_2008, haldane_biophysical_2014})
 assumed that a stationary distribution of mismatch classes is reached
in the evolution of isolated TFBS and thus an equilibrium solution, Eq~(\ref{equilibriumsoln}), is informative for binding sequence distributions.
In contrast, our results for average gain and loss times suggest that the evolution of an isolated TFBS is typically slow. 
To analyze this problem in a way that does not depend on arbitrary thresholds defining ``strong'' and ``weak'' binding classes $\mathcal{S}$ and $\mathcal{W}$, we first examined the evolution of the distribution $\boldsymbol{\psi}(k)$ over the mismatch classes as a function of time in Fig~\ref{Fig3}A. For typical parameter values it takes on the order of the inverse point mutation rate to reach the stationary distribution for populations that start off far away from it, even with strong selection. 

A systematic study of the convergence rates can be performed by computing the (absolute value of the) second eigenvalue, $|\lambda_2|$, of the transition rate matrix $\mathbf{R}$  from Eq~(\ref{transitionr}), and exploring how this depends on the biophysical parameters $n$ and $\epsilon$. Consistent with previous results, we observe large increases in convergence times as $n$ and $\epsilon$ increase. For example, an increase in the binding site length from $n=7$ to $n=11$ at baseline specificity of $\epsilon=2\,k_BT$ would result in a ten-fold increase in the convergence time. 

The intuitive reason behind the slow convergence rates is in the bimodal nature of the distribution $\boldsymbol{\psi}(k)$ on the thermodynamically motivated fitness landscape, similar to that reported by Lynch \& Hagner~\cite{lynch_2015}. One ``attractor'' is located around the fittest class ($k\sim0$, due to directional selection), while the
other is located around the redundancy-rich mismatch classes ($k\sim3/4n$). These two attractors are separated by a typically sharp fitness landscape, and the redundancy-rich attractor lacks selection gradients needed to support fast adaptation. The temporal evolution of the distribution $\boldsymbol{\psi}(k)$ from, e.g., a maladapted state, can thus be best understood as the probability weight ``switching'' from resting approximately within one attractor to the other one, while maintaining the bimodal shape throughout, rather than a gradual shift of a unimodal distribution from a maladapted initial value of $k$ to the value favored by selection.
This is especially true when $n$ gets larger: although adaptation within the functional sites can still happen, adaptation from the most random mismatch classes becomes extremely slow, even under strong selection (see Fig~\ref{FigS10}).

These results suggest that stationary distributions of isolated TFBS sequences may not be realizable on the timescales of speciation, which should be a cause of concern when stationarity is assumed without prior critical assessment. 
For example, applications assuming the stationary distribution might wrongly infer selection on regulatory DNA.

\subsection{Evolution of TF binding sites in longer sequences }

So far we have shown that the evolution of isolated TFBS is typically slow.
How do the results change if we consider TFBS evolution in a stretch of sequence $L$ bp in length, where $L\gg n$, e.g., within a promoter or enhancer?
Here we focus on \emph{de novo} evolution under strong directional selection for high gene expression, by simulating the process in the fixed state population framework. 
Compared to the isolated TFBS case, we need to make one further assumption: that the expression level of the selected gene is proportional to the summed TF occupancy on all sites within the regulatory region of length $L$ (see Models \& Methods for details). While this is the simplest choice, it is neither unique nor perhaps the most biologically plausible one, although limited experimental support exists for such additivity~\cite{giorgetti_noncooperative_2010}; it does, however, represent a tractable starting point when the interactions between individual TF binding sites are not strong and the contribution of each site is equal and of the same sign. 
To address the interactions, we look at the cooperative binding case in the following section.
In Supporting Information, we also discuss the competition of TFBSs for the strongest binding, and the ``nonphysical'' synergetic interaction by two strongest TFBSs.

We propose a simple analytical model for the time evolution of the number of strongly binding sites, $z(t)$, in the promoter, derived from isolated TFBS gain and loss rates, $\lambda_{\rm gain}$ and $\lambda_{\rm loss}$.  Assuming constant rates, one can write
\begin{equation}
\frac{d}{dt}z(t)=\lambda_{\rm gain}\Big(z_{\rm max}-z(t)\Big)-\lambda_{\rm loss}z(t)
\label{difeqn}
\end{equation}
where $z_{\rm max}$ is the maximum number of TFBS that can fit into the regulatory
sequence of length $L$ bp. If the sites can overlap, $z_{\rm max}=L-n+1$,
otherwise $z_{\rm max}\approx L/n$. The solution for Eq~(\ref{difeqn})  is 
\begin{equation}
z(t)=\Big(z_{\rm o}-\frac{B}{A}\Big)e^{-At}+\frac{B}{A}
\end{equation}
where $A=\big(\lambda_{\rm gain}+\lambda_{\rm loss}\big)$, $B=z_{\rm max}\lambda_{\rm gain}$
and $z_{\rm o}=z(t=0)$. Under strong positive selection, i.e. $Ns\gg n\log(2)/2$, the loss rate $\lambda_{\rm loss}$ can be ignored. 
If the distribution of the initial mismatch classes in the promoter is $\psi_{k}$, one can approximate $z_{\rm max}-z_{\rm o}=z_{\rm max}\,\sum_{k=k_{\mathcal{S}}+1}^{n}\psi_{k}$ to  obtain:
\begin{equation}
z(t)-z_{\rm o}=\big(1-e^{-\lambda_{\rm gain}t}\big)\, z_{\rm max}\,\sum_{k=k_{\mathcal{S}}+1}^{n}\psi_{k}.
\label{eq:Newnt}
\end{equation}
There are two limiting regimes in which we can examine the behavior of Eq~(\ref{eq:Newnt}). Over a short timescale, evolutionary dynamics will search over all possible positions, $z_{\rm max}=L-n+1$, to pull out the presites, since they are fastest to evolve into the strong binding class $\mathcal{S}$, i.e.:
\begin{equation}
\lambda_{\rm gain}\approx \lambda_{\rm gain}^{\rm presite}=\Big(\sum_{k\notin\mathcal{S}}\psi_{k}\Big)^{-1}\,\psi_{k_{\mathcal{S}}+1}/\langle t\rangle_{\mathcal{S}\leftarrow k_{\mathcal{S}}+1}
\label{eq:Newnt1}
\end{equation}

As the process unfolds and new sites are established, new TFBS will only be able to emerge at a smaller set of positions due to possible overlaps, so that $z_{\rm max}\approx L/n$. On the other hand, evolution from higher mismatch classes will also start to  contribute towards new sites:
\begin{equation}
\lambda_{\rm gain} \approx \lambda_{\rm gain}^{\rm all}=\Big(\sum_{k\notin\mathcal{S}}\psi_{k}\Big)^{-1}\,\sum_{k\notin\mathcal{S}}\psi_{k}/\langle t\rangle_{\mathcal{S}\leftarrow k}
\label{eq:Newnt2}
\end{equation}

\begin{figure}
\begin{centering}
\includegraphics[width=0.95\textwidth]{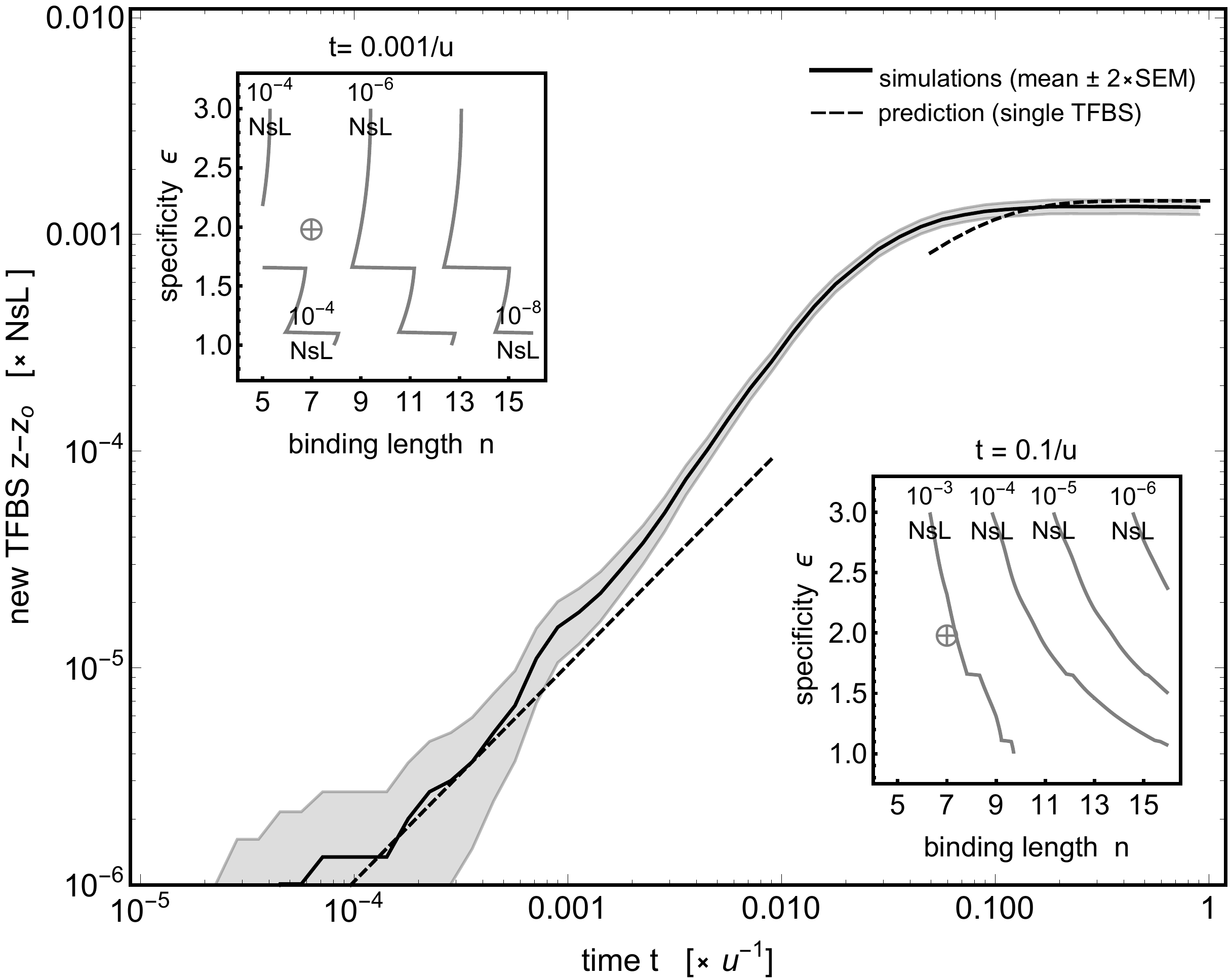}
\end{centering}
\caption{
\textbf{TF binding site evolution in a longer sequence of $L=30$ base pairs.}
The expected number of newly evolved TF binding sites with
length $n=7$ bp, under strong directional selection ($Ns=100$) and both point and indel mutations ($\theta=0.15$). Time is measured in inverse mutation rates; the number of newly evolved sites is scaled to the selection strength and the sequence length. 
1000 replicate simulations were performed with different initial sequences. Average number of sites shown by a solid black line; the gray band shows $\pm 2$ SEM (standard error of the mean) envelope. Dashed curves are analytical predictions based on single TFBS gain rates at an isolated DNA region, given by Eqs~(\ref{eq:Newnt},\ref{eq:Newnt1},\ref{eq:Newnt2}). Biophysical parameters used: $\epsilon=2$ $k_BT$, $\mu=4$ $k_BT$.
\textbf{Insets:}
Expected number of newly evolved sites
from a random sequence of length $L$ at $t=0.001 u^{-1}$ (left) and $t=0.1 u^{-1}$ (right) for different binding length and specificity
values, computed using the analytical predictions. 
Crosshairs denote the values used in the main panel.
\label{Fig4}}
\end{figure}

Fig~\ref{Fig4} shows how new TFBSs with length $n=7$ bp emerge 
over time in a promoter of $L=30$ bp in length. Consistent with the predictions of our simplified model, we can distinguish the early, intermediate, and late epochs.
In the early epoch, $t < 0.01 u^{-1}$, presites are localized among all possible locations and are established as binding sites. During this period, the growth in the expected number of new TFBSs is linear with time. The importance and predictive power of presites at early epoch remain even under different models of gene expression, including interaction between TFBSs (see Fig~\ref{FigS9}).
In the intermediate epoch, new binding sites accumulate at the rate that is slightly above that expected by establishment from presites alone, as the mutational neighborhood is explored further. In the late epoch, $t>0.1 u^{-1}$, initial sites in the immediate mutational vicinity have been exhausted, and established sites have constrained the number of positions where new sites can evolve from more distant initial sequences, leading to the saturation in the number of evolved TFBS. 

Using the simple analytical model, we explored in Fig~\ref{Fig4}B,C how the binding length $n$ and specificity $\epsilon$ affect the number of newly evolved TFBS. Increasing $n$ leads to a steep decrease in the number of expected sites, with a somewhat weaker dependence on $\epsilon$, especially at early times. 
Simulations at other values of biophysical and evolutionary parameters confirm the qualitative agreement between the analytical model and the simulation (Fig~\ref{FigS7}); given that the model is a simple heuristic, it cannot be expected to match the simulations in detail, yet it nevertheless seems to capture the gross features of evolutionary dynamics.
Together, these results show that at early times under strong selection, the number of newly evolved sites will grow linearly with time and proportional to $L$, before evolution from higher mismatch classes can contribute and ultimately before the sites start interacting, with a consequent slowdown in their evolution. Thus, evolution in longer regulatory regions ($L=10^2-10^3$ bp) could feasibly give rise to tens of binding sites  at $Ns=10^2-10^3$ within a realistic time frame $t\sim 0.001 u^{-1}$, if the sites are sufficiently short ($n\sim 7$ bp). Explaining the evolution of longer sites, e.g., $n>10-12$ bp, especially within short promoters found in prokaryotes, would likely necessitate invoking new mechanisms. 

\subsection{Ancient sites and cooperativity between TFs can accelerate binding site emergence}

Finally, we briefly examine two mechanisms that can further speed up the evolution of TF binding sites in longer sequences. 

The first possibility is that the sequence from which new TFBS evolve is
not truly random; as discussed previously, presites have a strong influence on the early accumulation of new binding sites. 
There are a number of mechanisms that could bias the initial sequence distribution towards presites: examples include transposable elements, DNA repeats, or CG content bias.
Here we consider an alternative mechanism that we refer to as the ``ancient TFBS scenario,'' in which a strong TFBS existed in the sequence in the ancient past, after which it decayed into a weak binding site, possibly due to the relaxation of selection (i.e., $Ns\sim 0$). 

As we demonstrated in the context of isolated sites, TFBS loss rates are slow and the remains of the binding site will linger in the sequence for a long time before decaying into the most redundancy rich mismatch classes. This biased initial distribution of mismatches $\boldsymbol \Psi$ in a sequence of length $L$ with a single ancient site can be captured by writing:
\begin{equation}
\boldsymbol \Psi = \frac{1}{L-n+1} \; \boldsymbol \psi (t') + \frac{L-n}{L-n+1} \; \boldsymbol \phi 
\label{eq:PresiteAncient}
\end{equation}
where $\boldsymbol \phi$ is the binomial distribution, Eq~(\ref{eq:binomialdistr}), characteristic of the random background, and $\boldsymbol \psi (t')$ is the distribution of mismatches due to the presence of the ancient site. Time $t'$ refers to the interval in which the isolated ancient TFBS has been decaying under relaxed selection, and the corresponding $\boldsymbol \psi (t')$ can be solved for using Eq~(\ref{eq:timesolutionR}). 

\begin{figure}
\begin{centering}
\includegraphics[width=\textwidth]{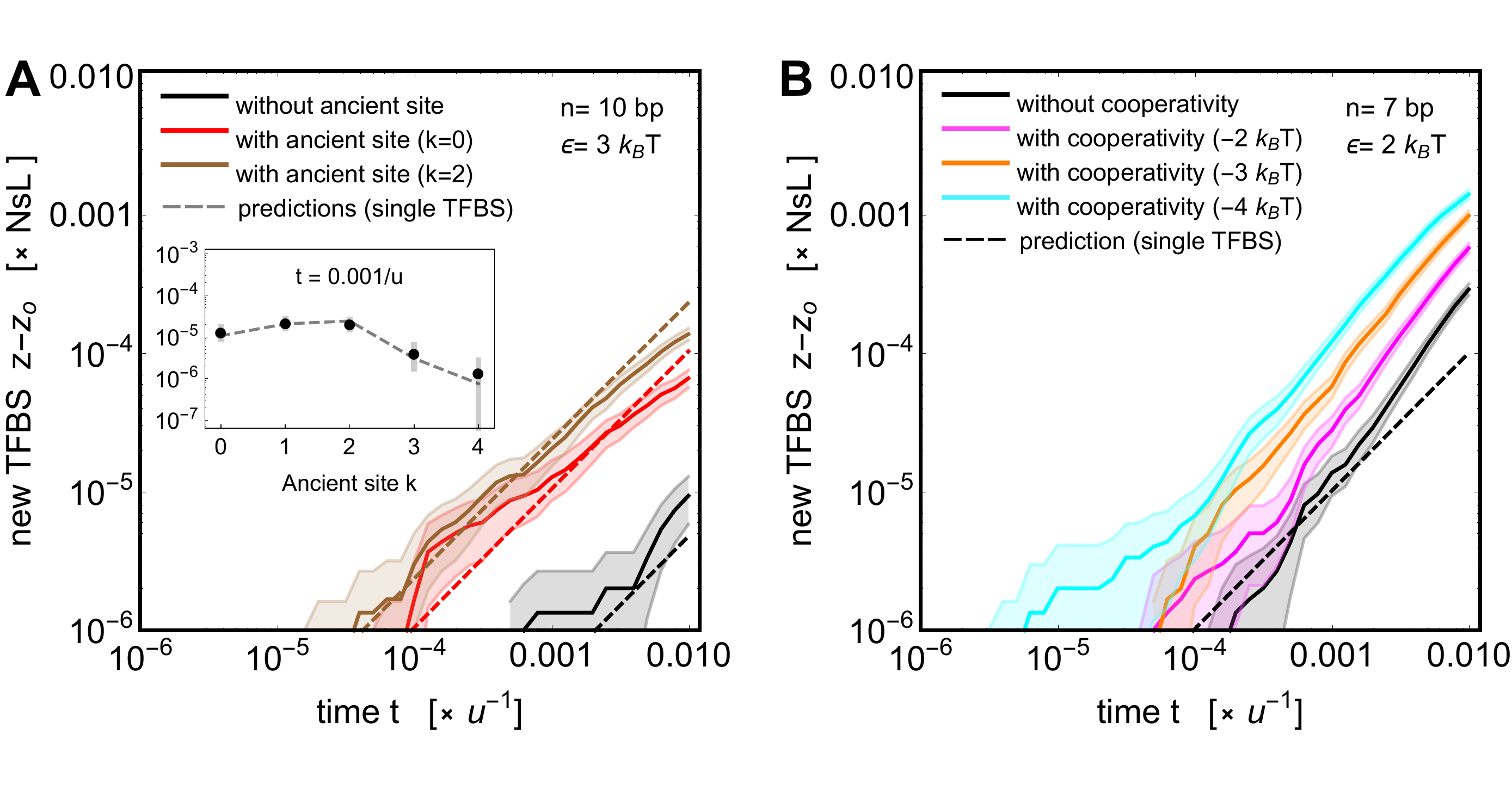}
\end{centering}
\caption{
\textbf{Ancient sites and cooperativity can accelerate the emergence of TF binding sites in longer regulatory sequences.}
{\bf A)} The expected number of newly evolved TFBS in the presence (red and brown) or absence (black) of an ancient site, for binding site length $n=10$ bp, and specificity, $\epsilon= 3\, k_BT$. In this example, the ancient site was a consensus site ($k=0$) or two mismatches away from it ($k=2$) that evolved under neutrality for $t'=0.1/u$ prior to starting this simulation. Dashed lines show the predictions of a simple analytical model, Eq~(\ref{eq:PresiteAncient}). The inset shows how the number of newly evolved TFBS at $t=0.001/u$ scales with the mismatch of the ancient site $k$ (plot markers: simulation means; error bars: two standard errors of the mean; dashed curve:  prediction). \textbf{B)} The expected number of newly evolved TFBS without (black) and  with cooperative interactions (for different cooperativity strengths, magenta: $E_c=-2 \, k_BT$, yellow: $E_c=-3 \, k_BT$, cyan: $E_c=-4 \, k_BT$, see Eq~(\ref{eq:TDmodelwcoop}) in the Models \& Methods and text) for binding site length $n=7$ bp, and specificity, $\epsilon= 2\, k_BT$. Both panels use $\mu=4$ $k_BT$, strong selection ($Ns=100$) and a combination of point and indel mutations ($\theta=0.15$), acting on a regulatory sequence of length $L=30$ bp. Thick solid lines show an average over 1000 simulation replicates, shading denotes $\pm 2$ SEM.
}
\label{Fig5}
\end{figure}

\begin{figure}
\begin{centering}
\includegraphics[width=\textwidth]{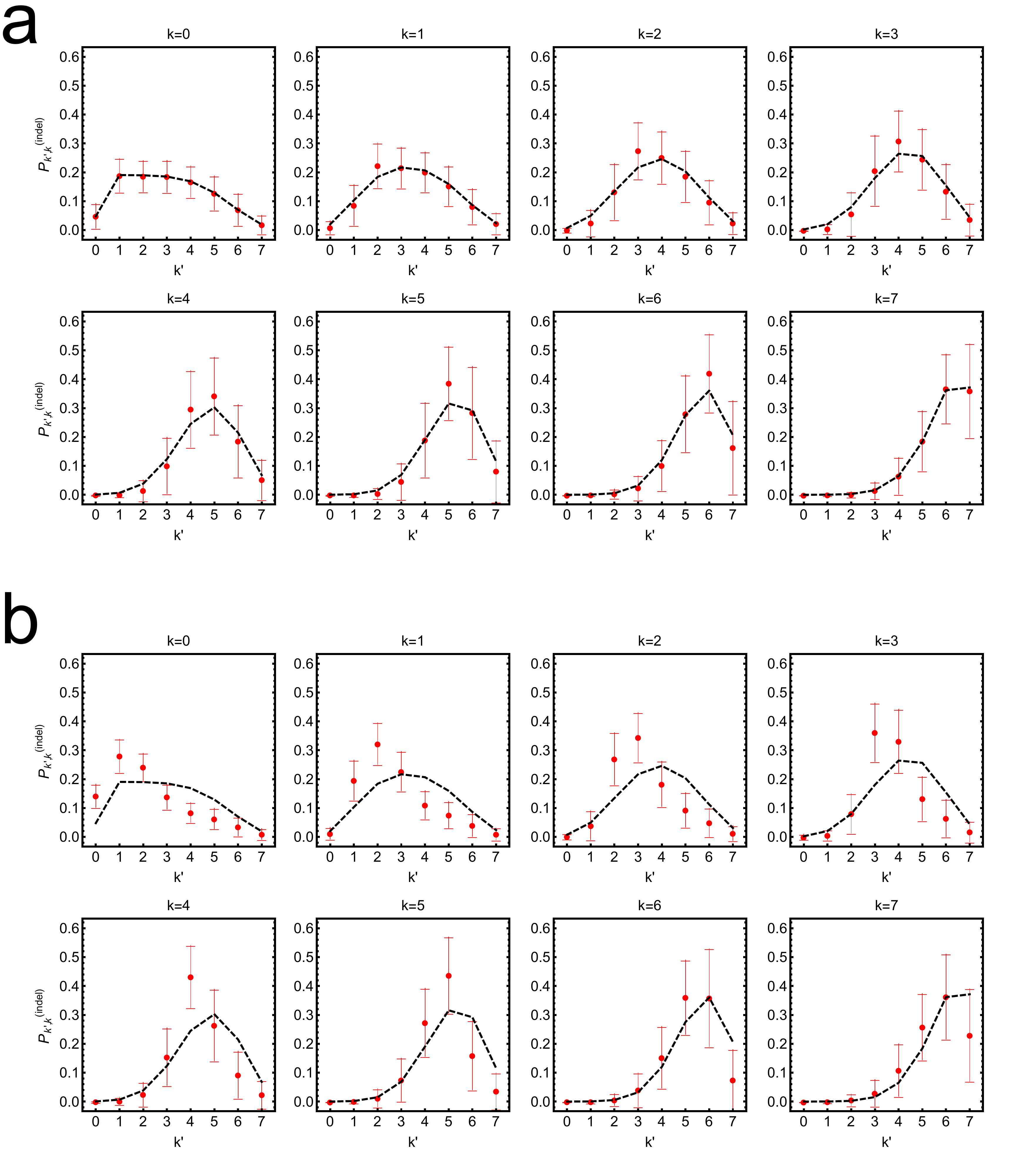}
\end{centering}
\caption{
\textbf{Indel mutations connect the mismatch genotype space differently from point mutations.}
\textbf{a)} Probability that a binding site with $k$ mismatches
mutates to $k'$ mismatches, for a single binding site of length $n=7$
bp, according to our indel mutation model in a fixed genomic window (see the Models \& Methods section).
Dashed curve = analytical prediction according to Eq~(\ref{eq:indel-mutation}).
Red points = mean $\pm 1$ std of $10^3$ replicate realizations
of the frequency distribution (for each replicate, $1$ consensus
sequence is created and $10^4$ mutations are simulated for each
$k$).
\textbf{b)} The same analysis as in a), but allowing for a flexible genomic window for alignment after insertion mutations. We pick the minimal mismatch case to asses the quality of our approximation. As expected, this creates a bias towards smaller mismatch classes, but suggests that our approximation is still reasonable.}
\label{FigS1}
\end{figure}

\begin{figure}
\begin{centering}
\includegraphics[width=5in]{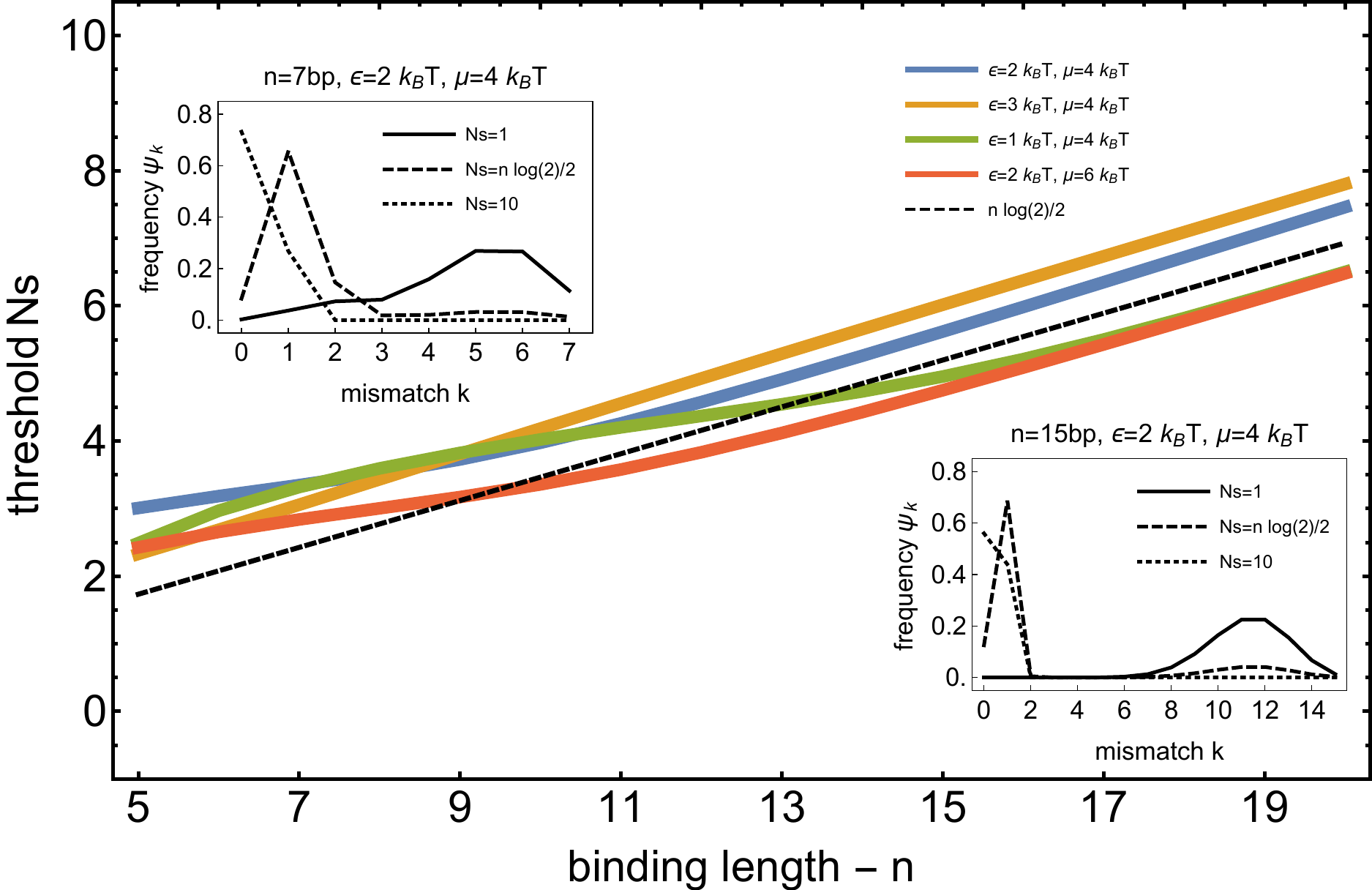}
\end{centering}
\caption{
\textbf{Threshold value of Ns for bimodality (i.e., threshold between strong and weak selection regimes)}. 
The value of $Ns$ at which $5\%$ of the probability weight in the stationary distribution is in non-strong mismatch classes, i.e. $k>k_\mathcal{S}$. For selection stronger than this threshold, the stationary distribution is concentrated at low $k$ (high fitness) classes and is practically unimodal. Different colors correspond to different biophysical parameters (see legend), analytical prediction $n \log (2)/2$ is in black (see the Models \& Methods section and Eq~(\ref{eq:thresholdNS})). Insets show examples of stationary distributions for different $Ns$ values for short and long binding sites.
}
\label{FigS2}
\end{figure}

Fig~\ref{Fig5}A shows that the ancient site scenario can enhance the 
number of newly evolved sites by resurrecting the ancient site, even after it has decayed for $t'=0.1u^{-1}$. Simulation results agree well with the simple analytical model using the biased initial sequence distribution of Eq~(\ref{eq:PresiteAncient}).
Importantly, such a mechanism is particularly effective for longer binding sites of high specificity, indicating that regulatory sequence reuse could be evolutionarily beneficial in this biophysical regime (see Fig~\ref{FigS8}).

Fig~\ref{Fig5}A and Fig~\ref{FigS8} also show the emergence of new sites when the ancient site was not a full consensus (preferred) sequence but differed from it by a certain number of mismatches. 
The results qualitatively agree with the case of perfect consensus. 
Importantly, this shows that the applicability of the ancient site scenario extends to cases where the ancient site belonged to a different TF (albeit with a preferred sequence similar to the studied TF), which has recently been reported to be a frequent phenomenon by Payne \& Wagner (2014)~\cite{payne_robustness_2014}, possibly due to evolution of TFs by duplication and divergence~\cite{weirauch_determination_2014}.

The second mechanism that we consider is the physical cooperativity between TFs: when one site is occupied, it is favorable for the nearby site to be occupied as well. 
We extended the thermodynamic model to incorporate cooperativity (see the Models \& Methods, Eq~(\ref{eq:TDmodelwcoop}) and Fig~\ref{Fig1}D). The genotype of a nearby site will then influence whether a given site acts as a strongly or weakly binding site.
The presence of a cooperative site acts as a local shift in the chemical potential, which changes the weak/strong threshold, so that an individually weak site can become a strongly binding site. 
Simulations using cooperative binding presented in Fig~\ref{Fig5}B illustrate how cooperativity can increase the speed of evolution. 
This is specifically effective for short binding sites of intermediate or low specificity, where a cooperative energy contribution can strongly influence the number of sites in the strong binding class (see Fig~\ref{FigS8}).

\section{Discussion}

In this study, we aimed at a better theoretical understanding of which biophysical and population genetic factors influence the fast evolution of TFBSs in gene regulatory DNA, making sequence specific TF binding a plausible mechanism for the evolution of gene regulation and for generating phenotypic diversity.
Following Berg et al. (2004)~\cite{berg_adaptive_2004}, we combined a biophysical model for TF binding with a simple population genetic model for the rate of sequence evolution. 
The key assumptions are that binding probability is determined by a thermodynamic equilibrium; that fitness depends linearly on binding probability; and that populations are typically homogeneous in genotype, and so evolve by substitution of single point and short insertion/deletion (indel) mutations. 
Remarkably, the biophysical and the evolutionary models take the same mathematical form: in the biophysical model, binding probability depends on the binding energy, relative to thermal fluctuations, $\beta E$, whilst in the evolutionary model, the chance that a mutation fixes depends on its selective advantage, relative to random sampling drift, $Ns$.

For single TFBS evolution, we calculated the average transition time between genotypes, the inverse being  a measure for the speed of the evolution. 
Our results indicate that TFBS evolution is typically slow unless selection is very strong. It is important to emphasize that gaining a TFBS by point mutations under neutral evolution is very unlikely, contrasting with the belief in the current literature (e.g., \cite{wittkopp_evolution_2013, villar_evolution_2014}). 
This is mainly due to Stone \& Wray's argument that functional sites could readily be found by a random walk~\cite{stone_rapid_2001}; however, their argument assumed that individuals follow independent random walks, which grossly overestimates the rate of evolution (see MacArthur \& Brookfield~\cite{macarthur_expected_2004}). 
Indeed, fast rates of gaining a single TFBS require not only strong selection but also initial sequences in the mutational neighborhood of the functional sites. 
Especially, ``presites," i.e. sequences 1 bp away from threshold sequences, can be crucial since they can evolve to functional sites by single mutations. 
Indel mutations can increase the rate of gaining a single TFBS from distant sequences, since they connect the genotype space extensively, but their effect is limited under realistic indel mutation rates~\cite{cartwright_problems_2009, chen_variation_2009}. 
Future studies should consider the updates in estimates of indel mutation rates, since they are currently not as precise as point mutation rates, although we do not expect big qualitative departures from our results.

Considering the evolution of a single TFBS from random sequence, we showed that biophysical parameters, binding length and specificity, are constrained for realistic evolutionary gain rates from the most redundant mismatch classes. The rates  drop exponentially with binding length, making TF whose binding length exceeds $10-12$ bp difficult to evolve from random sites, at least under the point and indel mutation mechanisms considered here. As a consequence of the biophysical fitness landscape, binding specificity and length show an inverse relation for the same magnitude of the gain rate from the most redundant mismatch class. Such an inverse relation is observed in position weight matrices of TFs collected from different databases for both eukaryotic and prokaryotic organisms, by Stewart \& Plotkin (2012)~\cite{stewart_why_2012}. In the same study, they reproduce this observation using a simple model which assumes that a trade-off between the selective advantage of binding to target sites, versus the selective disadvantage of binding to non-target sequence. Their model assumes a stationary distribution, and that sites are functional if they are mismatched at no more than one base. It would be interesting to explore a broader range of models that account for the dynamical coevolution between  transcription factor binding specificity, its length, and its binding sites~\cite{lynch_2015}. One idea can be to combine the evolutionary dynamical constraints (against large binding length and high specificity, which we show here) with simple physical constraints of TF dilution in non-target DNA (against short binding length and low specificity, again in an inverse relation~\cite{gerland_physical_2002}).

For a single TF binding site, the stationary distribution for the mismatch with the consensus binding sequence depends on the binding energy, but also on the sequence entropy -- that is, the number of sequences at different distances from the consensus.  
Typically, the distribution is bimodal: either the site is functional, and is maintained by selection, or it is non-functional, and evolves almost neutrally. 
We show that it may take an extremely long time for the stationary distribution to be reached. 
Functional sites are unlikely to be lost if selection is strong (i.e., $Ns\gg1$), whilst function is unlikely to evolve from a random sequence by neutral evolution, even if predicted under stationarity assumption. 
Therefore, typical rapid convergence to stationary distribution should be considered with caution in theoretical studies.

We showed that the dynamics of TFBS evolution in longer DNA sequences can be understood from the dynamics of single TFBS. 
The rate of evolution of new binding sites will be accelerated in proportion to the length of the promoter/enhancer sequence in which that can be functional; however, because this increase is linear in promoter/enhancer length, it will have a weaker influence than the exponential effect changes in specificity or length of binding site. 
Especially the earlier dynamics (relevant for speciation timescales) are determined by the availability of presite biased sequences. 
Any process that allowed selection to pick up more distant sequences or that increased presite ratio among non-functional sites would accelerate adaptation from ``virgin'' sequences.

A key factor for an enrichment in presite ratio may arise through variation in GC content or through simple sequence repeats (especially if the preferred sequence has some repetitive or palindromic structure). 
In this study, we showed that it may also arise from ancient sites, i.e. sites that were functional in earlier evolutionary history and decayed into nonfunctional classes in evolution. 
Since loss of function is slow (comparable to the neutral mutation rate once selection becomes ineffective), this is plausible for sites that are under intermittent selection, or where there is a shift to binding by a new TF with similar preferred sequence~\cite{payne_robustness_2014, weirauch_determination_2014}. 
This effect of the earlier evolution can be especially important for long binding TFs as convergence to a truly randomized sequence distribution requires much longer times. 
MacArthur and Brookfield~\cite{macarthur_expected_2004} showed that real promoter sequences may acquire functional sites more quickly than random sequence, but it is not clear whether that is due to a different general composition, or to the ghosts of previous selection. 
New studies are required to test our enriched presite-biased sequence hypothesis, especially for orthologous regions where functional TFBS is observed in sister populations or species. 
In a recent study, Villar \textit{et al.} (2015)~\cite{villar_enhancer_2015} provide evidence that enhancer DNA sequence structure is older than other DNA portions, suggesting the reuse of such regions in evolution, plausibly by gaining and losing TFBSs in repetitive manner. 
Nourmohammad \& Lassig (2011)~\cite{nourmohammad_formation_2011} showed evidence suggesting that local duplication of sequences followed by point mutations played important role in binding site evolution in Drosophila species (but surprisingly, not in yeast species).
Another interesting option  would be the existence of ``mobile'' presites or their fragments, e.g., as sequences embedded into transposable elements that could be inserted before the gene under selection for high expression~\cite{feschotte_transposable_2008}. Presites can be considered as concrete examples of cryptic sequences \cite{rajon_compensatory_2013}, potential source of future diversity and evolvability. 
We believe that understanding the effects of presites would contribute to the predictability of genetic adaptations regarding gene regulation, especially in important medical applications such as antibiotic resistance or virus evolution.

We also showed that the evolution of a functional binding site in longer DNA can be accelerated by cooperativity between adjacent transcription factors. 
When a TF occupies a co-binding site, sufficient transcriptional activity can be achieved from sequences of larger mismatch classes, an effect similar to a local increase in TF concentration. This mechanism permits faster evolution towards strongly binding sequences, and seems most effective for short TFBS where it creates a selection gradient already in the redundancy rich mismatch classes.
Cooperative physical interactions might allow the evolution of binding occupancy and thus expression without large underlying sequence changes, which might be a reason for the  observed  weak correlation between sequence and binding evolution at certain regulatory regions.
Importantly, TFBS clustering in eukaryotic enhancers can be a consequence of the fast evolution with cooperativity, as also supported by a recent empirical study~\cite{stefflova_cooperativity_2013}. 

Our theoretical framework is relevant more broadly for understanding the evolution of gene regulatory architecture. Since the speed of TFBS evolution from random sequences is proportional to $NsL$, our results suggest that population size $N$ and the length of regulatory sequences $L$ can compensate for each other in terms of the rate of adaptation.
This is exactly what is observed: eukaryotes typically have longer regulatory DNA regions but small population sizes, while prokaryotes evolve TFBS within shorter regulatory sequence fragments but have large population sizes. Similarly, prokaryotes might have achieved longer TF binding lengths $n$,  as large population size allowed them to overcome the exponential decrease in the gain rates with increasing $n$. If relevant, these observations would suggest that an important innovation in eukaryotic gene regulation must have been the ability of the transcriptional machinery to integrate the simultaneous occupancy of many low-specificity transcription factors bound over hundreds of basepairs of regulatory sequence, a process for which we currently have no good biophysical model.

\section{Acknowledgments}
We thank Magdalena Steinr\"uck, Georg Rieckh, Ferran Palero and Ziya Kalay for helpful comments. NB acknowledges support from ERC Advanced Grant 250152 ``Selection and Information.''

\section{References}
%
%
%
%
%



\pagebreak
\section{Supporting Information}

\subsection{Other fitness models for comparison \& for interacting TFBSs}

\subsubsection{Power-law decaying fitness models for comparison:} 
\begin{figure}
\begin{centering}
\includegraphics[width=5.5in]{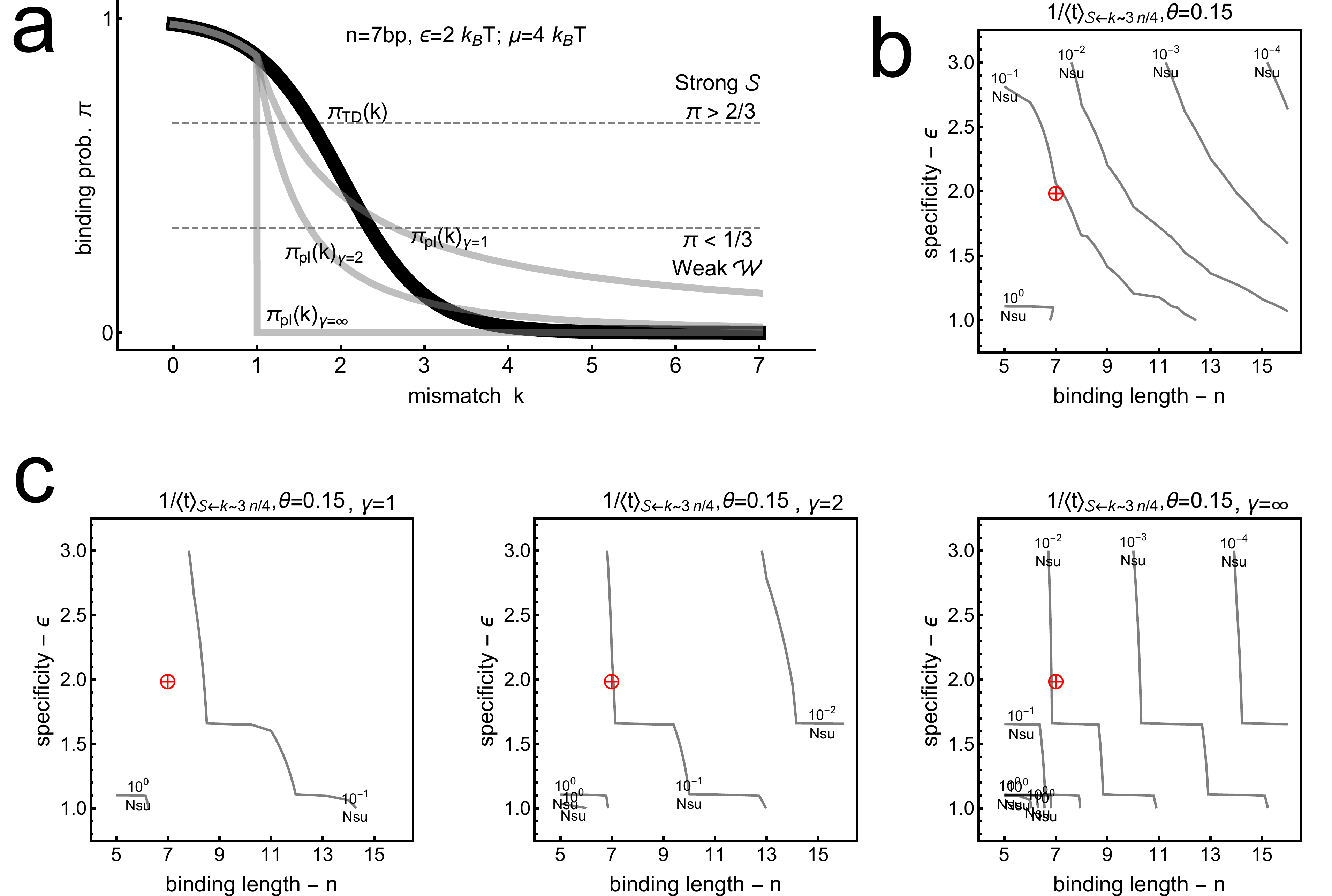}
\end{centering}
\caption{
\textbf{Single TFBS gain rates in modified fitness landscapes with a power-law tail.}
The thermodynamic fitness landscape has been modified to have a power-law decaying tail of exponent $\gamma$ for  $k>k_\mathcal{S}$, as in Eq~(\ref{eq:modifiedfitness}) in SI text. We tested $\gamma=1$, $2$ and $\infty$ corresponding to smooth, intermediate and step-like decay. Plot conventions are the same as in Fig~\ref{Fig2}C.
\textbf{b)} Isolated TFBS gain rate from the most redundant mismatch class for the thermodynamic model, replotted from Fig~\ref{Fig2}C for reference.
\textbf{c)} Plots analogous to b) using modified fitness landscapes defined by the power-law exponent $\gamma$. Gain rates are higher for small $\gamma=1$ and lower for the step landscape ($\gamma=\infty$), relative to the reference.
}
\label{FigS3}
\end{figure}

\begin{figure}
\begin{centering}
\includegraphics[width=4in]{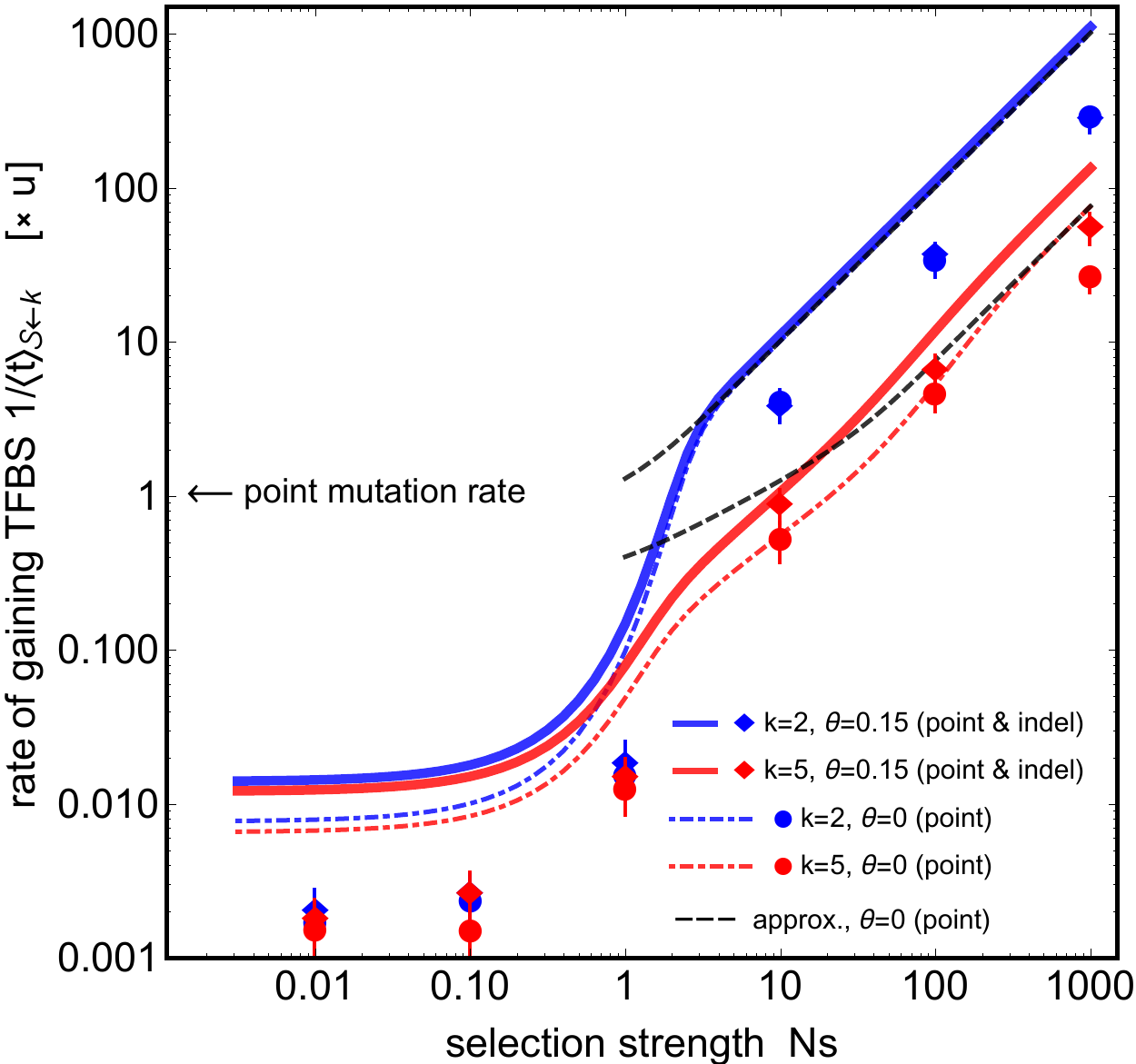}
\end{centering}
\caption{
\textbf{The effect of polymorphisms on the single TFBS gain rate at higher mutation rates.}
Wright-Fisher simulation results (point markers, error bars = 2 standard errors of the mean) at $4Nu=0.1$, in comparison to the fixed state model (continuous curves). Plot conventions are the same as in Fig~\ref{Fig2}. Biophysical parameters used: $n=7$, $\epsilon=2$ $k_{B}T$, $\mu=4\; k_BT$. Polymorphisms generally decrease TFBS gain rates.
}
\label{FigS4}
\end{figure}

\begin{figure}
\begin{centering}
\includegraphics[width=4in]{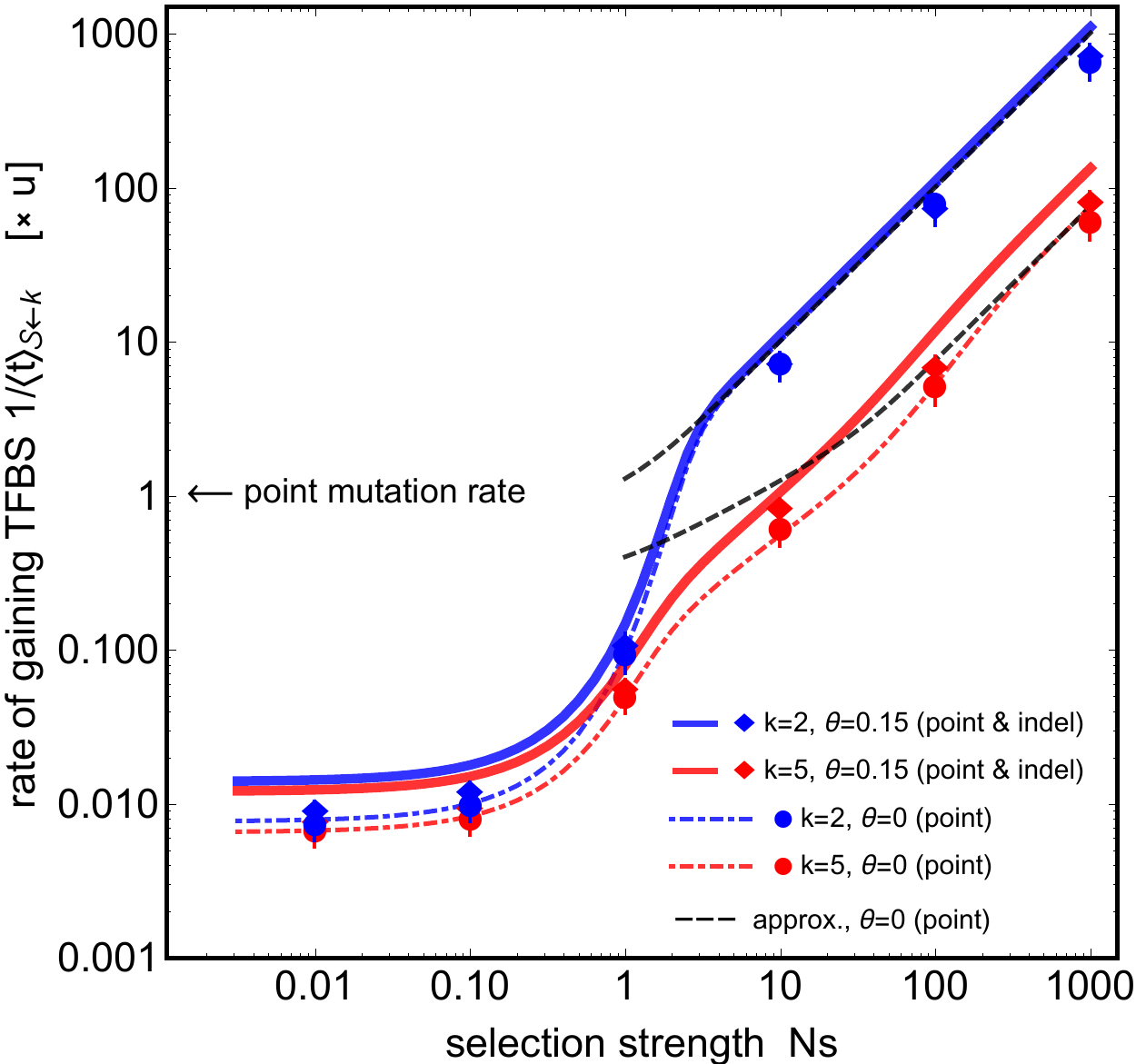}
\end{centering}
\caption{
\textbf{Relaxing the mismatch assumption.}
Fig~\ref{Fig2}, but using energy matrices whose nonzero entries are gaussian random variables $\varepsilon_i$, such that $\langle \varepsilon_i\rangle=\epsilon=2k_BT$ and $\sigma_\varepsilon=0.5k_BT$; $n=7$, $\mu=4k_BT$. The analytical results under the equal mismatch assumption are shown in continuous lines.
}
\label{FigS5}
\end{figure}

\begin{figure}
\begin{centering}
\includegraphics[width=\textwidth]{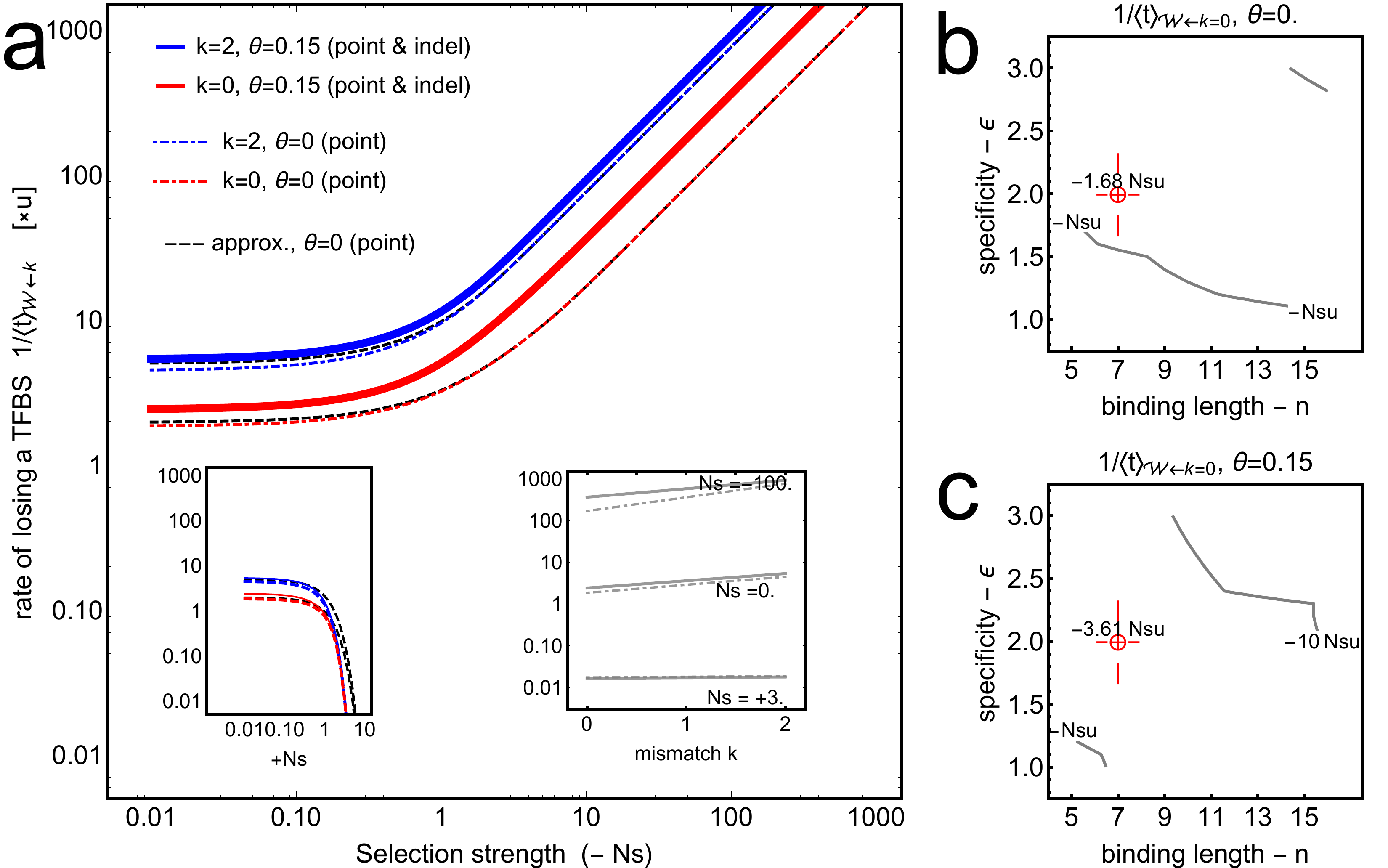}
\end{centering}
\caption{
\textbf{Single TF binding site loss rates at an isolated DNA region.}
The dependence of the loss rate, $1/\langle t\rangle_{\mathcal W \leftarrow k}$ shown in units of point mutation rate, from sequences in different initial mismatch classes $k$ (blue: $k=2$, red: $k=0$),
as a function of negative selection strength. Results with point mutations only ($\theta=0$) are shown by dashed line; with admixture of indel mutations ($\theta=0.15$) by a solid line. For strong selection, $|Ns|\gg 1$, the rates scale with $2|Ns|nu$, which is captured well by the ``shortest path'' approximation (black dashed lines in the main figure) of Eq~(\ref{eq:AverageTimeShortestPathGain}). The biophysical parameters are:
site length $n=7$ bp; binding specificity $\epsilon=2$ $k_B T$; chemical
potential $\mu=4$ $k_{B}T$. Left inset: $Ns$-scaling with positive selection.
Right inset: gain rates as a function of the initial mismatch class $k$ for different $Ns$.
\textbf{b, c)} Loss rates from the consensus sequence ($k=0$) under strong negative selection, without (b) and with (c) indel mutations supplementing point mutations. Red crosshairs denote the cases depicted in panel a). Contour lines show constant loss rates in units of $Ns\,u$ as a function of biophysical parameters $n$ and $\epsilon$. 
}
\label{FigS6}
\end{figure}

\begin{figure}
\begin{centering}
\includegraphics[width=\textwidth]{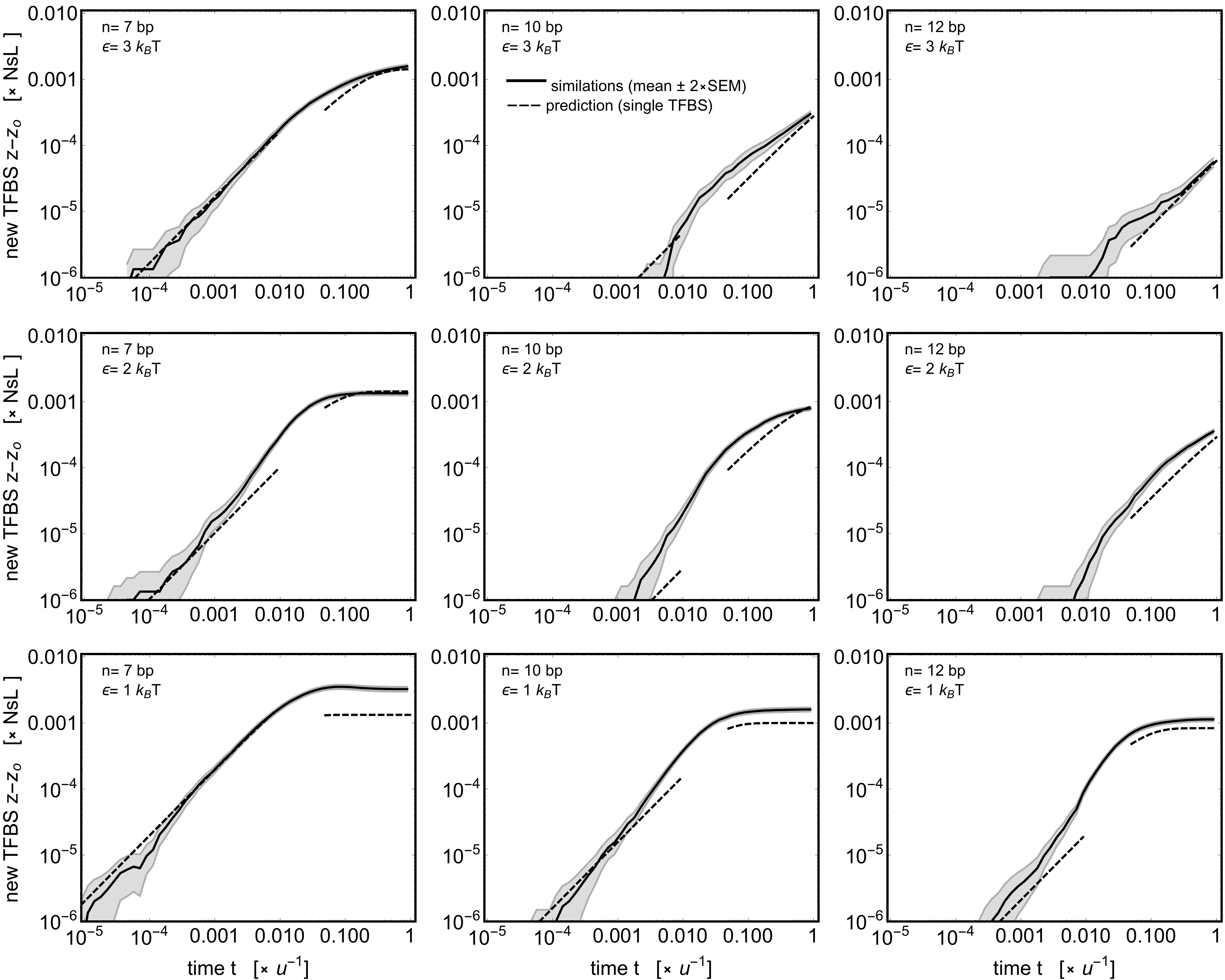}
\end{centering}
\caption{
\textbf{TFBS evolution in longer sequences}.
Example simulations (black solid line) and analytic predictions based on single TFBS gain/loss rates (black dashed line), for different binding length $n$ and specificity $\epsilon$. Details are identical to Fig.~\ref{Fig4}.
}
\label{FigS7}
\end{figure}

In order to understand the importance of the thermodynamically-motivated sigmoid shape for the binding probability, we compare our results to those obtained with power-law functions that decay with exponent $\gamma$ (note that $\gamma=\infty$ corresponds to a step-like fitness landscape), formally defined as
\begin{equation}
\quad\pi_{{\rm pl}}(k)=
\begin{cases}
\pi_{\rm TD}(k) & k\leq k_{\mathcal S} \\
\big(k_{\mathcal S} / k \big)^\gamma \; \pi_{\rm TD} (k_{\mathcal S}) & k > k_{\mathcal S}
\end{cases}.
\label{eq:modifiedfitness}
\end{equation}
Fig~\ref{FigS3} shows that the power-law exponent is a major determinant of the gain rates, suggesting that a biophysically realistic fitness landscape is crucial for the quantitative understanding of TFBS evolution.

\subsubsection{Fitness models of interacting TFBSs in larger regulatory sequence:}
\begin{figure}
\begin{centering}
\includegraphics[width=\textwidth]{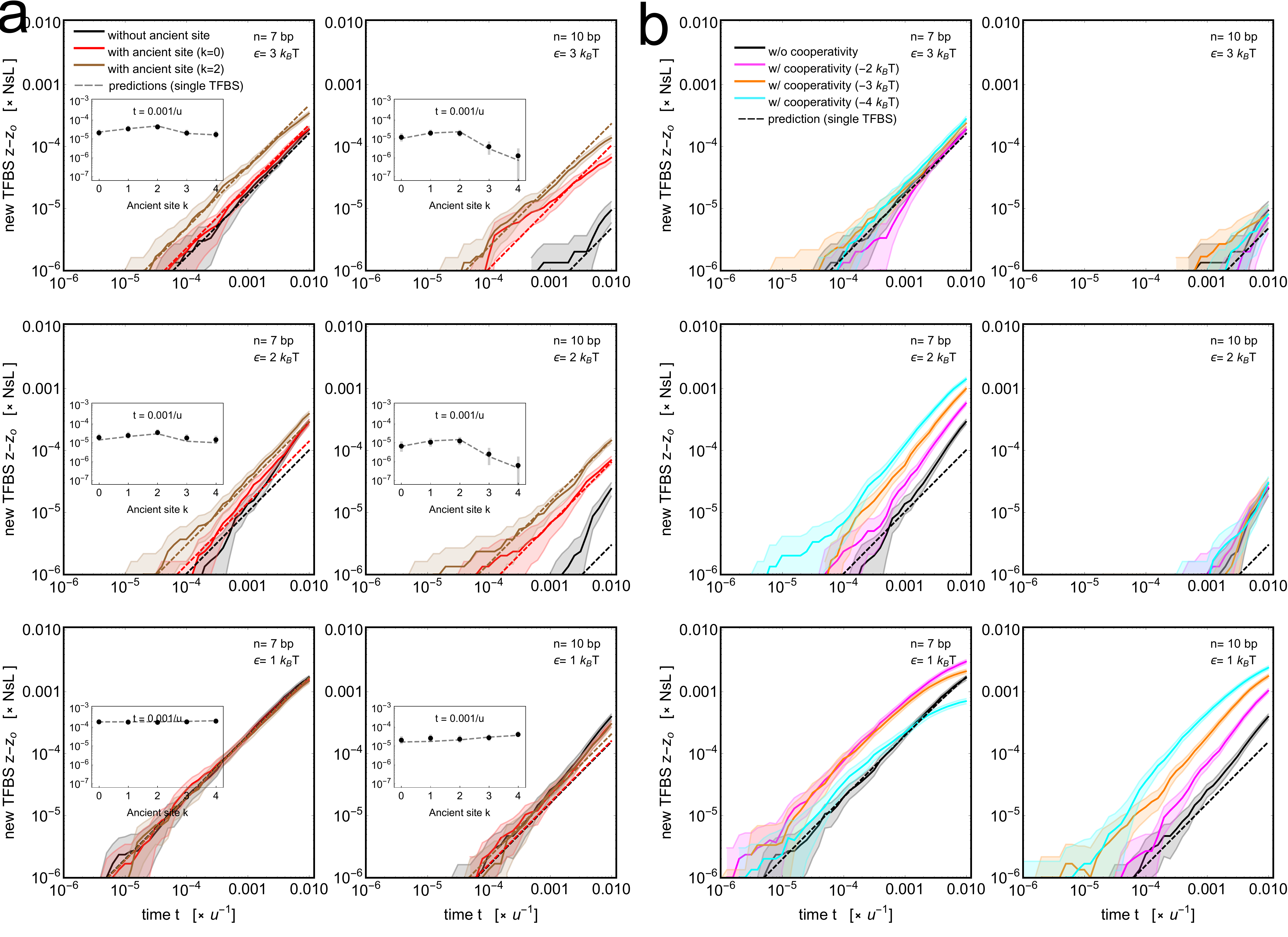}
\end{centering}
\caption{
\textbf{The effect of ancient sites (a) and cooperativity (b) for different binding lengths and specificities}.
Simulations of TFBS evolution in longer sequences (colored lines) and analytic predictions based on single TFBS gain and loss rates (dashed black lines), analogous to Fig.~\ref{Fig5}. Different panels show different choices of TFBS binding length $n$ and specificity $\epsilon$. Ancient sites specifically facilitate the emergence of longer sites of high specificity, whereas cooperativity specifically facilitates the emergence of shorter sites of intermediate or low specificity.
}
\label{FigS8}
\end{figure}

In addition to physical cooperativity between nearby TFs on promoter/enhancers (see the Models \& Methods, Fig~\ref{Fig5} and  Fig~\ref{FigS8}), here we also consider two other models.
The first additional model assumes that the binding occupancy of the strongest binding site in the regulatory sequence is the proxy for the gene expression level and the fitness, i.e.
\begin{equation}
f(\boldsymbol \sigma) = s \,\rm{MAX}\{ \pi^{(i)}(\boldsymbol \sigma) \}.
\label{1StrongTFBS}
\end{equation}
Note that different TFBSs interact with each other to compete for the strongest binding  within a promoter or an enhancer.

The second additional model addresses synergistic interaction between the two strongest-binding TFBS, located anywhere in the regulatory sequence. This example is a simplified version of a  biophysical model where  TFs, binding anywhere in a regulatory region, compete for the occupancy of that region with a nucleosome (for a more elaborative modeling framework, see  Mirny (2010)~\cite{mirny_nucleosome-mediated_2010}). We call this type of interaction between two TFs ``non-physical'' because TFs don't interact directly; their interaction is effectively mediated by some other biophysical process. 
The probability of the joint occupancy of the two TFs at promoter or enhancer can be used as the proxy for gene expression level and the fitness, i.e. 
\begin{equation}
f(\boldsymbol \sigma) = s \,\frac{e^{-\beta(\epsilon (k_1+k_2) - 2\mu)} } {1 + e^{-\beta ( \epsilon k_1 - \mu)} + e^{-\beta (\epsilon k_2 - \mu)} + e^{-\beta(\epsilon (k_1 + k_2) - 2\mu)}},
\label{SynergeticTDmodel}
\end{equation}
where $k_1$ and $k_2$ correspond to the genotypes of two TFBSs with the smallest mismatches in the regulatory sequence.

\begin{figure}
\begin{centering}
\includegraphics[width=0.9\textwidth]{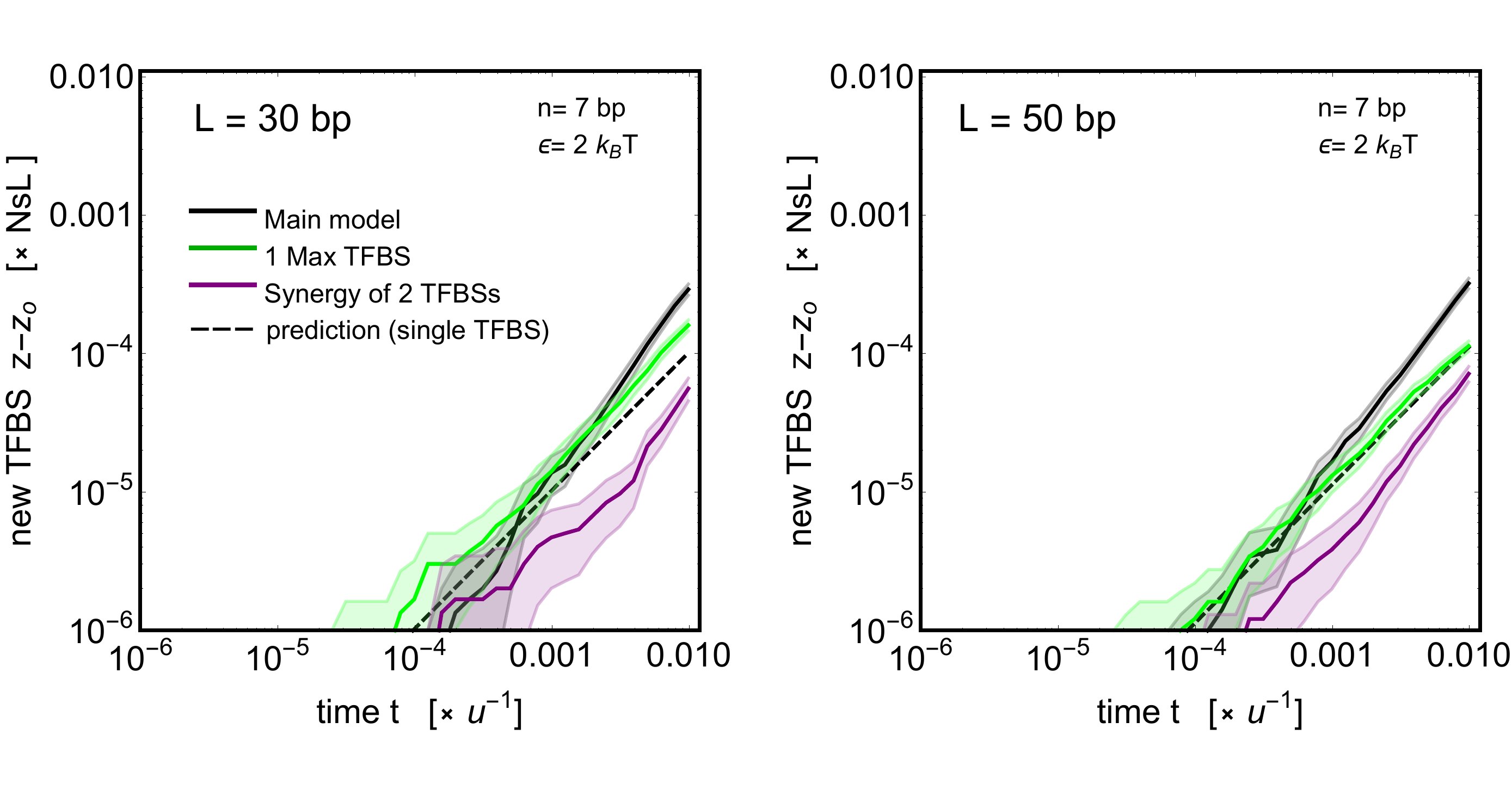}
\end{centering}
\caption{
\textbf{Fitness models of interacting TFBSs.}
The expected number of newly evolved TFBS for binding site length $n=7$ bp, specificity $\epsilon= 2\, k_BT$, and chemical potential $\mu= 4\, k_BT$ are shown  for different fitness models. The solid black curve is the non-interacting model used in the main text (dashed curve: theoretical prediction).  The green curve stands for the model of Eq~(\ref{1StrongTFBS}) in SI text, where only the strongest binding site in the regulatory sequence determines gene expression. The purple curve stands for the model of Eq~(\ref{SynergeticTDmodel}) in SI text, where two strongest TFBS synergistically determine the gene expression level. Shading denotes $\pm 2$ SEM. The simulations use regulatory sequences of length $L=30$ bp (left) and $L=50$ bp (right).
}
\label{FigS9}
\end{figure}

\begin{figure}
\begin{centering}
\includegraphics[width=0.9\textwidth]{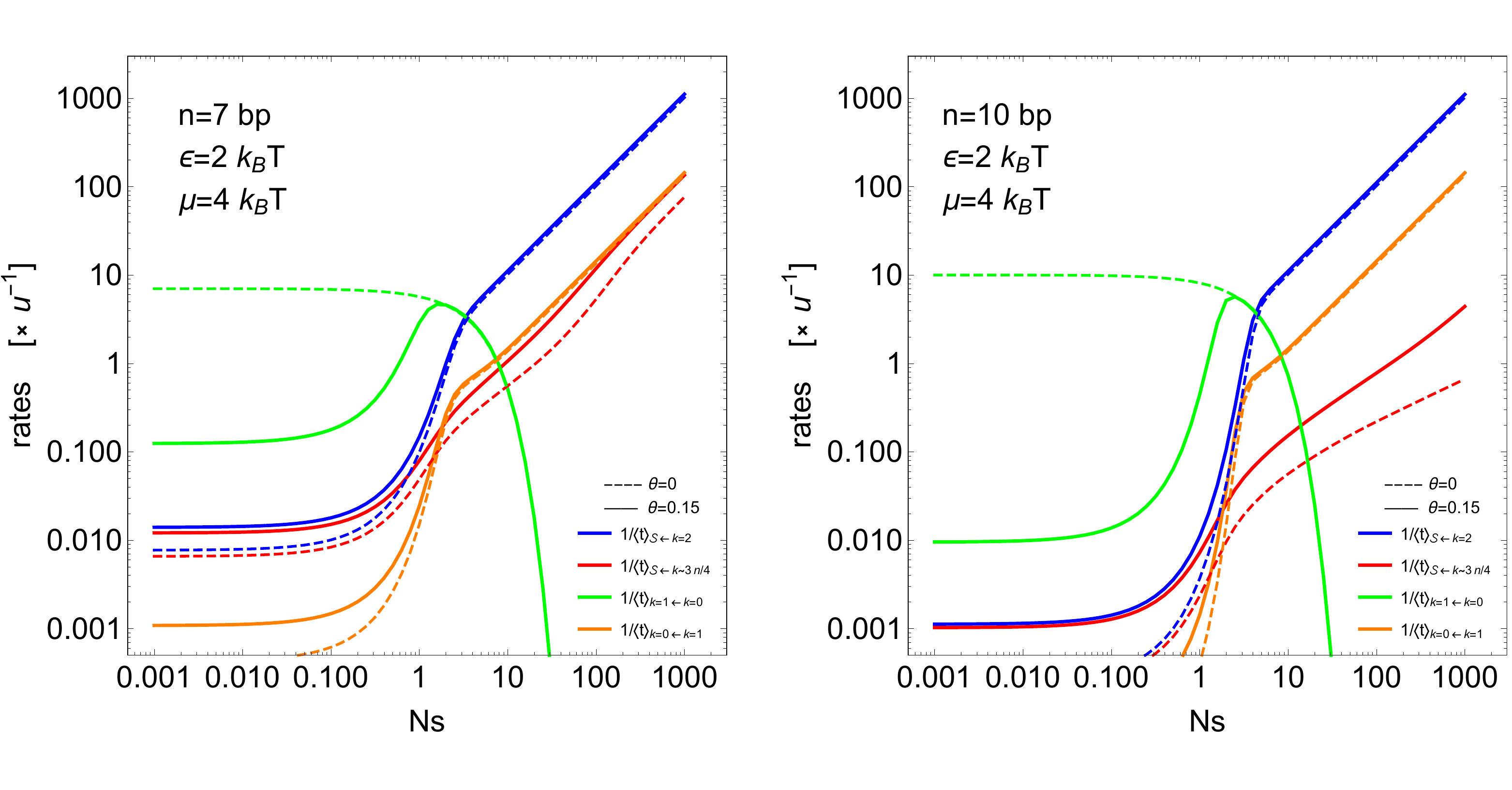}
\end{centering}
\caption{
\textbf{Comparison rates of TFBS gain rates and sequence turnover rates within functional TFBSs}
Average first hitting times to particular mismatch $k_j$ state can be calculated with a minor modification to Eq~(\ref{eq:averagetimematrix}) by replacing $\mathcal S$ with $k_j$. The figures compare the rates of evolution of TFBS within the functional sites (i.e. $1/\langle t \rangle_{k=0 \leftarrow k=1}$ and $1/\langle t \rangle_{k=1 \leftarrow k=0}$). Plot conventions are the same as in Fig~\ref{Fig2}-A. Biophysical parameters used: $n=7$ bp (left), $n=10$ bp (right) $\epsilon=2$ $k_{B}T$, $\mu=4\; k_BT$. It shows that for weak selection, the rates to evolve from $k=0$ to $k=1$ can be relatively faster. Also, although adaptation from random sites slows down with increasing $n$, we see that the adaptation rate to evolve from $k=1$ to $k=0$ can stay high.
}
\label{FigS10}
\end{figure}

Do these models yield different result for the emergence of strong binding sites from random sequences at early evolutionary times ($\sim$ speciation time scales), in comparison to our main model, where the sum of binding occupancies is used as a proxy for gene expression level [Eq(\ref{GeneExpressionModel}) in the main text]? For typical biophysical parameters (binding lenght: $n=7$ bp, binding specificity: $\epsilon=2$ $k_{B}T$ and chemical potential: $\mu=4$ $k_{B}T$), we show in Fig~\ref{FigS9} that these modified models do not differ extensively from results of our main model.

\end{document}